\documentclass[fleqn,usenatbib]{mnras}

\usepackage{newtxtext,newtxmath}

\usepackage[T1]{fontenc}

\DeclareRobustCommand{\VAN}[3]{#2}
\let\VANthebibliography\thebibliography
\def\thebibliography{\DeclareRobustCommand{\VAN}[3]{##3}\VANthebibliography}

\newcommand\sersic{S\'ersic}
\newcommand\umg{\mu_{\text{molgas}}}
\newcommand\mmg{M_{\text{molgas}}}

\newcommand\scotwo{S_{\text{CO(2-1)}}}
\newcommand\lcotwo{L_{\text{CO(2-1)}}^{'}}
\newcommand\lcoone{L_{\text{CO(1-0)}}^{'}}

\usepackage{indentfirst}
\usepackage{appendix}
\usepackage{makecell}
\usepackage{enumerate}
\usepackage{graphicx}
\usepackage{pdflscape}
\usepackage{float}
\usepackage{lineno}
\usepackage{scrextend}
\deffootnote{0.25em}{0em}{\thefootnotemark\quad}
\DeclareUnicodeCharacter{2212}{-}
\DeclareUnicodeCharacter{00B4}{'}
\DeclareUnicodeCharacter{00A8}{"}
\usepackage{amsmath}

\title[ALMA dual quasars]{Rich and diverse molecular gas environments of closely-separated dual quasars viewed by ALMA}

\author[S. Tang et al.]{
Shenli Tang,$^{1,2,3}$\thanks{E-mail: st1c23@soton.ac.uk}
John D. Silverman,$^{2,4,5,6}$
Zhaoxuan Liu,$^{2,4}$
Manda Banerji,$^{3}$
Tomoko Suzuki,$^{4}$\newauthor
Seiji Fujimoto,$^{7}$
Andy Goulding,$^{8}$
Masatoshi Imanishi,$^{9}$
Toshihiro Kawaguchi,$^{10}$
Connor Bottrell,$^{11,2}$\newauthor
Tilman Hartwig,$^{12}$
Knud Jahnke,$^{13}$
Masafusa Onoue,$^{2,14}$
Malte Schramm,$^{15}$
and Yoshihiro Ueda$^{16}$
\\
$^{1}$Department of Physics, University of Tokyo, Tokyo 113-0033, Japan\\
$^{2}$Kavli Institute for the Physics and Mathematics of the Universe (WPI), The University of Tokyo, Kashiwa, Chiba 277-8583, Japan\\
$^{3}$School of Physics \& Astronomy, University of Southampton, Highfield Campus, Southampton SO17 1BJ, UK\\
$^{4}$Department of Astronomy, School of Science, The University of Tokyo, 7-3-1 Hongo, Bunkyo, Tokyo 113-0033, Japan\\
$^{5}$Center for Data-Driven Discovery, Kavli IPMU (WPI), UTIAS, The University of Tokyo, Kashiwa, Chiba 277-8583, Japan\\
$^{6}$Center for Astrophysical Sciences, Department of Physics \& Astronomy, Johns Hopkins University, Baltimore, MD 21218, USA\\
$^{7}$Department of Astronomy, The University of Texas at Austin, Austin, TX, USA\\
$^{8}$Department of Astrophysics, Princeton University, Princeton, NJ08544, USA\\
$^{9}$National Astronomical Observatory of Japan, 2-21-1 Osawa, Mitaka, Tokyo 181-8588, Japan\\
$^{10}$Department of Economics, Management and Information Science, Onomichi City University, Hisayamada 1600-2, Onomichi, Hiroshima 722-8506, Japan\\
$^{11}$International Centre for Radio Astronomy Research, University of Western Australia, 35 Stirling Hwy, Crawley, WA 6009, Australia\\
$^{12}$Institute for Physics of Intelligence, The University of Tokyo, Tokyo, Japan\\
$^{13}$Max-Planck-Institut für Astronomie, Königstuhl 17, 69117 Heidelberg, Germany\\
$^{14}$Kavli Institute for Astronomy and Astrophysics, Peking University, Beijing 100871, China\\
$^{15}$Graduate school of Science and Engineering, Saitama Univ., 255 Shimo-Okubo, Sakura-ku, Saitama City, Saitama 338-8570, Japan\\
$^{16}$Department of Astronomy, Kyoto University, Kitashirakawa-Oiwake-cho, Sakyo-ku, Kyoto 606-8502, Japan
}

\date{Accepted XXX. Received YYY; in original form ZZZ}

\pubyear{\the\year{}}

\begin{document}
\label{firstpage}
\pagerange{\pageref{firstpage}--\pageref{lastpage}}
\maketitle

\begin{abstract}
We present a study of the molecular gas in five closely-spaced ($R_{\perp}<20$ kpc) dual quasars ($L_{\rm bol}\gtrsim10^{44}~\mathrm{erg~s}^{-1}$) at redshifts $0.4<z<0.8$ with the Atacama Large Millimeter/submillimeter Array. The dual quasar phase represents a distinctive stage during the interaction between two galaxies for investigating quasar fueling and feedback effects on the gas reservoir. The dual quasars were selected from the Sloan Digital Sky Survey and Subaru/Hyper Suprime-Cam Subaru Strategic Program, with confirmatory spectroscopic validation. Based on the detection of the CO J=2--1 emission line with Band 4, we derived key properties including CO luminosities, line widths, and molecular gas masses for these systems. Among the ten quasars of the five pairs, eight have line detections exceeding $5\sigma$. The detected sources prominently harbor substantial molecular gas reservoirs, with molecular gas masses ($\mmg$) between $10^{9.6-10.5}~\mathrm{M_{\odot}}$, and molecular gas-to-stellar mass ratios ($\umg$) spanning $18-97\%$. The overall $\umg$ of these dual quasars agrees with that of inactive star-forming main-sequence galaxies at comparable redshifts, indicating no clear evidence of quenching. However, intriguing features in each individual system show possible evidence of AGN feedback, matter transfer, and compaction processes.
\end{abstract}




\section{Introduction}
\begin{figure*}
\centering
\includegraphics[width=.9\linewidth]{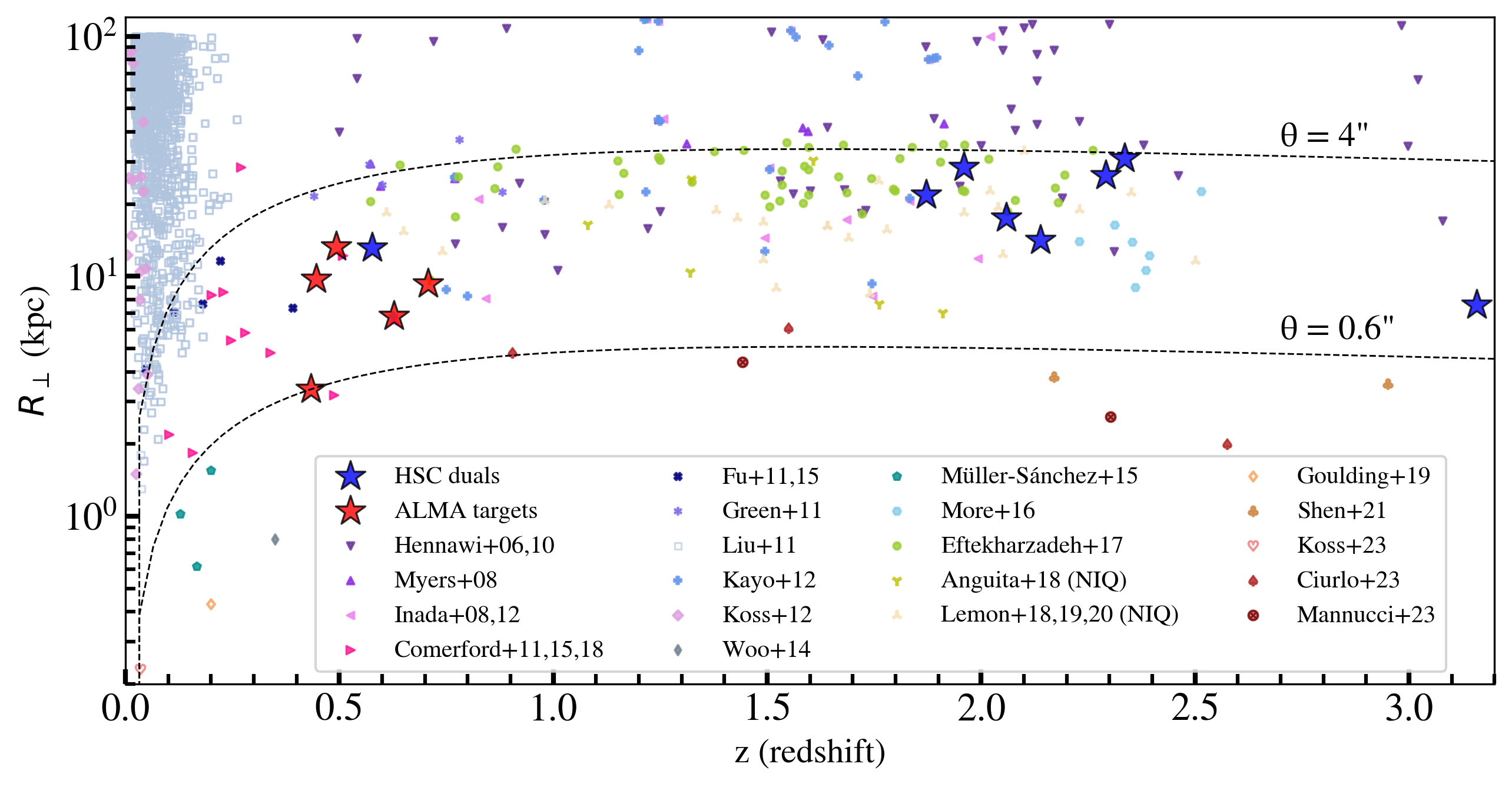}
\caption{Dual quasar compilation with projected physical distance ($R_{\perp}$ in kpc) versus redshift. Dual quasars confirmed through our HSC project with the five ALMA targets in red while the rest are in blue (Tang et al. in preparation). Spectroscopically-confirmed dual quasars from the literature are shown \citep[][]{hennawi2006binary,hennawi2010binary,myers2008quasar,inada2008sloan,inada2012sloan,green2011multiwavelength,liu2011active,fu2011kiloparsec,fu2015binary,comerford2011chandra,comerford2015merger,comerford2018origin,kayo2012very,koss2012understanding,woo2014sub,muller2015origin,more2016sdss,eftekharzadeh2017clustering,anguita2018strong,lemon2018gravitationally,lemon2019gravitationally,goulding2019discovery,lemon2020strong,shen2021hidden,koss2023ugc,ciurlo2023new,mannucci2023gmp}. The dashed curves mark the selection region between $0.6\arcsec-4\arcsec$ of our HSC project. In particular, the ALMA targets are between $0.6\arcsec-2.2\arcsec$.
\label{fig:separation}}
\end{figure*}

Quasar activity often intertwines with the hierarchical growth of galaxies. When galaxies collide, the resulting massive interaction can fuel star formation and potentially accelerate the growth of their central supermassive black holes (SMBHs) \citep[e.g.,][]{sanders1996luminous,hopkins2006unified}. Such interactions might be responsible for the observed correlations between the physical properties of SMBHs and their host galaxies \citep[e.g.,][]{di2005energy,di2008direct}. If both merging galaxies host an SMBH, it could lead to a phase where both SMBHs are simultaneously active, forming a dual quasar. The discovery of dual quasars, initially observed with separations spanning several arcseconds, dates back to \cite{owen1985vla}, and gained momentum with the Sloan Digital Sky Survey (SDSS) \citep{hennawi2006binary,hennawi2010binary}. However, detecting dual quasars with separations $\lesssim$ 20 kpc beyond the local Universe has posed a challenge due to resolution constraints, i.e.,  both sources must be resolved to confirm the separate nuclei. Figure~\ref{fig:separation} shows a compilation of dual quasars validated through optical spectroscopy, X-ray, or radio emission \citep[also see][]{chen2022varstrometry}. This includes Nearly Identical Quasars (NIQs) being reported in lens surveys, which are defined as quasar pairs with similar spectra but lacking photometric detections of a lens galaxy \citep{anguita2018strong,lemon2018gravitationally,lemon2019gravitationally,lemon2020strong}. High redshift dual quasars have been reported up to z=5.66 \citep{yue2021candidate} and z=6.05 \citep{matsuoka2024discovery}, and the spatial separations have been reported down to 430 pc \citep{goulding2019discovery} and 230 pc \citep{koss2023ugc}. This diversity in the redshifts and separations of dual quasars illustrates their existence across various merger stages of galaxies throughout cosmic time.
\par
Dual quasar systems have motivated simulation efforts to understand their prevalence and detailed dynamics within galactic interactions. Hydrodynamic cosmological simulations have calculated the dual fraction (the proportion of dual quasars among quasars) to range from a few thousandths to a few percent \citep{van2012observability,steinborn2016origin,silverman2020dual}. Some studies further propose a positive evolution with redshift \citep{rosas2019abundances,volonteri2016cosmic,volonteri2022dual}. In parallel, numerical simulations have delved into the evolving physical properties of dual quasars and their host galaxies as the merger progresses \citep{capelo2015growth,capelo2017survey}. Consequently, the quest for more dual quasars at close separations and subsequent investigation using multi-wavelength greatly advanced holds immense significance. Such endeavors aim to constrain these simulations and unravel the underlying physics governing the co-evolutionary processes \citep[see review by][]{de2019quest}.
\par
However, the properties of the host galaxies of dual quasars are still poorly studied due to the limited samples and the difficulties of disentangling the quasar and host light with spatial resolution issues, especially beyond the local universe. One approach is the 2D modeling of deep images. For example, using the Hubble Space Telescope (HST), \cite{chen2023close} examined the dual quasar SDSS J0749+2255 (at $z=2.17$, separated by 3.8 kpc) and found that it resides within massive compact disk-dominated galaxies displaying tidal features. This discovery serves as direct evidence of ongoing interactions in dual quasars, which may further evolve into compact gravitationally bound binary SMBHs. More recently, the James Webb Space Telescope (JWST) has shown the power of revealing the 3D structure of this system \citep{ishikawa2024vodka,chen2024vodka}, and also confirming the nature of similar candidates in various environments \citep{maiolino2023jades,perna2023ultradense,li2024jwst,ubler2024ga}.
\par
In comparison to space-based imaging, ground-based telescopes offer distinct advantages in efficiency and cost-effectiveness for conducting systematic surveys. Since our first paper \citep{silverman2020dual}, we have been carried out a search for dual quasars in the Subaru Hyper Suprime-Cam Subaru Strategic Program (HSC/SSP) footprint with close separation ($0.6\arcsec-4\arcsec$, shown by the dashed curves in Figure~\ref{fig:separation}). Candidate selection was based on morphology and color followed by comprehensive multi-wavelength spectroscopic analysis to confirm their nature and probe their physical characteristics. In \cite{silverman2020dual}, we spectroscopically confirmed three dual quasars using Keck/LRIS. Subsequently, \cite{tang2021optical} presented three additional confirmed dual quasars identified via Gemini-GMOS and Subaru/FOCAS. We estimated their black hole and host properties via optical spectroscopy and imaging. To date, we have only studied the stellar contents of their host galaxies, leaving the gas and dust properties unknown. 
\par
Extensive investigations of molecular gas have been conducted across various galaxy populations including single quasar host galaxies. Using the 30-meter IRAM telescope, \cite{xia2012molecular} observed 19 infrared ultraluminous quasars at $z<0.4$, in which they detected CO J=1--0 emission line in 17 sources, revealing a molecular gas content comparable to ULIRGs ($\mmg \sim 10^{9-10}~\mathrm{M_{\odot}}$). Similar findings were reported in \cite{krips2012co,husemann2017integral}. Using the Atacama Large Millimeter/submillimeter Array (ALMA), \cite{shangguan2020alma} studied CO J=2--1 emission in 23 Palomar-Green (PG) quasars at $z<0.1$, reporting a 91\% detection rate and a mean $\mmg$ value of $10^{9.20\pm0.13}\mathrm{M_{\odot}}$ for their quasar host galaxies, consistent with inactive galaxies of similar stellar mass. The lack of correlations between the quasar properties and the global CO properties are also reported in $z<0.5$ PG quasars \citep{molina2023lack} and $z<0.2$ type 2 quasars \citep{molyneux2024quasar}. On the other hand, \cite{izumi2020circumnuclear} observed 4 quasars and 4 matched inactive star-forming galaxies (SFGs) at $z<0.06$, and found higher molecular gas surface densities in SFGs than the quasars at scales less than 500 pc. Therefore, the low-z results tend to support a ``local" (e.g., circumnuclear-scale) impact of AGN feedback on its molecular gas environment. ALMA also greatly advanced studies of quasars during the Epoch of Reionization \citep{wang2013star,decarli2017rapidly,jones2017dynamical,venemans2017compact,willott2017wide,shao2017gas,izumi2018subaru,nguyen2020alma,yue2021alma,walter2022alma} and at cosmic noon  \citep{banerji2017discovery,banerji2021resolving,schulze2019no,scholtz2023evidence}. Interestingly, these studies tend to prefer a relatively-depleted gas environment of quasar hosts. Such a variation might be the results of the rising luminosities of the quasars at earlier universe \citep{shen2020bolometric}, so that higher-z quasars are likely to have more efficient feedback to quench their gaseous environments \citep{valentini2020impact}. The variations may also be attributed to higher obscuration rates and higher merger rates of high-redshift quasars \citep{treister2010major}. Therefore, fair comparison needs to be made under a well-constrained framework considering all these factors.
\par

Back to the dual quasars observed in this work, they are well-defined mergers with both spectroscopic and photometric evidences. Four among the five selected pairs are type1-type1 pairs, and one is a type1-type1.5 pair. The redshifts of the samples are between 0.4 and 0.8, bridging the gap between the local Universe and samples at cosmic noon. These facts make these five dual quasars an ideal pilot sample set to study the impact of quasar feedback on its molecular gas environment under a well-constrained framework. In the literature, there are no adequate observations of molecular gas in samples of dual quasars at this redshift range. In this work, we extend our investigation of five dual quasars by observing the CO J=2--1 emission using ALMA Band 4 to assess the molecular gas content of their host galaxies. This study aims to answer two main questions: (1) With two quasars being simultaneously activated, is the molecular gas component of dual quasars more depleted than the single quasars and inactive SFGs? (2) Are there any preferred spatial distributions and kinematic structures of the molecular gas among each pair?
\par
The paper is structured as follows: 
Section~\ref{sec:methods} details our sample selection, observation setup, and the procedure for data reduction.
In Section~\ref{sec:results}, we present the overall measurements concerning the observed CO properties and the inferred gas properties.
In Section~\ref{sec:individuals}, we dive into the details, and highlight the intriguing features of each pair. 
In Section~\ref{sec:discussion}, we compare the CO properties of our dual quasars with single quasars and inactive galaxies in the literature, aiming to answer whether these systems are quenched. Furthermore, we discuss implications for the connections between galaxy mergers and quasar activities based on our results.
Throughout this work, we adopt a $\Lambda$CDM cosmology with parameters $\Omega_{\Lambda}=0.7$, $\Omega_{\mathrm{m}}=0.3$, and $H_{0}=70 \mathrm{~km} \mathrm{~s}^{-1} \mathrm{Mpc}^{-1}$.

\section{Methods} \label{sec:methods}
\subsection{Sample selection} \label{subsec:selection}
Our selection of dual quasars originates from the SDSS DR14 quasar catalog \citep{paris2018sloan}, captured within the Subaru HSC/SSP footprint \citep{aihara2019second}. Using the 2D image modeling tool \textsc{GaLight} \citep{ding2022galight}, we selected quasars accompanied by another nearby point source with a separation of $0.6-4\arcsec$. These quasars typically span redshifts from 0 to 4, with bolometric luminosities typically exceeding $10^{44}~\mathrm{erg/s}$.
\par
The validation process involved spectroscopic observations to confirm the nature of the companion sources. Here, confirmation of a dual quasar requires the detection of a broad emission line in both sources, with the line center being offset by at most 2000 km/s following \cite{hennawi2006binary}. To date, we have confirmed 13 dual quasars (Tang et al. in preparation), as denoted by the star marks in Figure~\ref{fig:separation}. Among the five dual quasars targeted for ALMA observation in this study, highlighted in red, four belong to unobscured (type1-type1) quasar pairs, while one constitutes a partly obscured-unobscured (type1-type1.5) pair. The angular separations of these five dual quasars span $0.6-2.2\arcsec$, translating to projected physical separations ranging from $3.9-13.3$ kpc. This sample enables investigations of the molecular gas content and dynamics during a phase approximately 200 million years prior to the final coalescence of a galaxy merger, as expected from numerical simulations. \citep{capelo2015growth,capelo2017survey}.
\par

\begin{table*}
\caption{ALMA observation setups for the dual quasars. Columns (2)-(4): optical position and redshift of the SDSS quasar in the pair. See Table \ref{tab:optical} for reanalyzed values on both sources. (5) Sensitivity of the emission line in 10 km/s width. (7) Minimum velocity resolution. (8) Precipitable water vapor. (10) Total integration time in seconds.} 
\label{tab:obs_setup}
\begin{tabular}{cccccccccccc}
\hline
Name & RA & Dec & z & Line sens. & Beam size & $R_{v}$ & PWV & Humidity & $t_{\text{int}}$ & Date\\
(SDSS J2000) & (deg) & (deg) & & (mJy/beam) & (arcsec$^2$) & (km/s) & (mm) & (\%) & (s) & (yy-mm-dd)\\
(1) & (2) & (3) & (4) & (5) & (6) & (7) & (8) & (9) & (10) & (11)\\
\hline
084710.40-001302.6 & 131.79336 & 0.21741 & 0.6269 & 0.595 & $0.87\times0.73$ & 8.287 & 0.42 & 7.87 & 1,603 & 22-06-10\\
121405.12+010205.1 & 183.52136 & 1.03478 & 0.4927 & 0.664 & $0.84\times0.79$ & 7.611 & 2.04 & 28.35 & 1,300 & 22-01-07\\
141637.44+003352.2 & 214.15602 & 0.56452 & 0.4336 & 0.559 & $0.32\times0.24$ & 7.331 & 1.16 & 7.49 & 2,480 & 21-11-28\\
220906.91+004543.9 & 332.27881 & 0.76219 & 0.4461 & 0.679 & $0.89\times0.66$ & 7.374 & 1.92 & 8.90 & 1,360 & 21-12-30\\
233713.66+005610.8 & 354.30695 & 0.93634 & 0.7078 & 0.574 & $0.51\times0.48$ & 8.003 & 4.75 & 30.63 & 1,814 & 21-12-05\\
\hline
\end{tabular}
\end{table*}

\subsection{Observations and data reduction} \label{subsec:observation}
We conducted the observations during ALMA Cycle 8, spanning from November 2021 to June 2022 (Project ID: 2021.1.01233.S, PI: Tang). These observations entailed mapping the CO J=2--1 emission using a configuration involving 41--48 12-meter antennas equipped with the Band 4 receiver. Each spectral window within the 1.875 GHz bandpass was divided into 480 spectral channels (with an averaging factor of 8), resulting in a spectral resolution of 3904 kHz. This resolution corresponds to an approximate velocity resolution of $\sim$8 km/s at the observed frequency. The combined frequency coverage, encompassing 4 spectral windows across two sidebands, amounted to 7.5 GHz. The observational setup, parameters, and weather conditions are recorded in Table \ref{tab:obs_setup}.
\par
Our data reduction process uses CASA version 6.4.1 \citep{bean2022casa} with the delivered Measurement Set (MS) files. Each MS file, derived from the raw data, undergoes calibration through the execution of the auxiliary ``scriptForPI.py" file, thus is also known as the ``calibrated visibility". For each individual source, a comprehensive suite of calibrators, including atmosphere, bandpass, flux, phase, pointing, and water vapor radiometers (WVR), is utilized in this process. Detailed information about this calibration procedure can be accessed in the ALMA Science Pipeline User's Guide \footnote{https://almascience.nrao.edu/processing/science-pipeline}, and the QA reports in the ALMA project repository.
\par
Based on the MS file, our initial step involves splitting the visibility into emission line and continuum windows through the \texttt{split} task. In the emission line window, we extract the source spectrum and estimate the FWHM of the line using the spectral profile tool within \texttt{casaviewer}. Then the other parts of the spectra excluding $\pm 5 \times \mathrm{FWHM}$ of the emission line, including the continuum windows, are stacked to capture the continuum emission of the source. Under our observation setups, the continuum of all of our five sources turn out to be negligible.
\par
We further split the emission line visibility using a window centered on the line with a width of $2 \times \mathrm{FWHM}$. We then use the \texttt{tclean} task to generate a dedicated emission-line image. Within \texttt{tclean}, the \texttt{Multiscale} deconvolver \citep{cornwell2008multiscale} is used to capture potential extended structures within the data. To process both continuum and cube data, the \texttt{specmode} is configured as multifrequency synthesis (\texttt{mfs}) for continuum and \texttt{cube} mode for cube data, respectively. We extend the usage of \texttt{mfs} mode for the emission line following the method in \cite{akins2022alma}. This approach yields an average line intensity map measured in units of Jy/beam, denoted as the moment 0 (M0) map within this study. In the results of this paper, we set the robustness parameter (R) for Briggs weighting \citep{briggs1995high} to 2 (i.e., natural weighting) to maximize sensitivity, because the main aim of this work is to estimate the total amount of molecular gas. The peak fluxes of the sources with R=0.5 weighting differ from R=2 by $\lesssim$ 15\%.
\par
It is worth noting that the M0 map could vary depending on the center and width chosen for \texttt{tclean}. For the cases where we have double detections, we always make two versions of the M0 map, with each centered on one source. All the follow-up measurements are also made accordingly (detailed in Section \ref{sec:individuals}).
\par

On the M0 map, we use the interactive 2D fitting tool within casaviewer to pre-estimate the spatial center, size, and peak flux of the sources. Subsequently, these pre-estimated values serve as initial parameters for the subsequent \texttt{imfit} task, facilitating the estimation of integrated line flux measured in Jy. Utilizing this line flux, we then calculate the line luminosity according to \cite{solomon1992warm}:
\begin{equation} \label{eq:L_line}
L_{\text {line }}^{\prime}=3.25 \times 10^7 \times S_{\text {line }} \Delta v \frac{D_L^2}{(1+z)^3 \nu_{obs}^2} \mathrm{~K} \mathrm{~km} \mathrm{~s}^{-1} \mathrm{pc}^2
\end{equation}
where $S_{\text {line }}$ is the measured CO J=2--1 flux in Jy. $\Delta v$ is $2 \times \mathrm{FWHM}$ of the emission line, defines the width applied to the \texttt{mfs} mode of \texttt{tclean} in our specific analysis. $D_L$ is the luminosity distance in Mpc, and $\nu_{\text{obs}}^2$ is the observed frequency of the emission line.
\par
In handling the cube data, we uniformly set the channel width to 10 km/s across all sources. Within each channel, the standard deviation typically hovers around $\sim 10^{-4}$ Jy/beam. Taking the channels within $\sim 2 \times \mathrm{FWHM}$ of the emission line and retaining stacked pixels above $\sim 3\sigma$, we generate the moment 1 (M1) and moment 2 (M2) maps through \texttt{immoments}. The exact frequency region and the pixel threshold are detailed for each source in Section \ref{sec:individuals}. From the cube data, we extract the 1D spectrum of each source within an elliptical aperture. Then the same aperture is used to extract the noise spectrum from the background. In practice, we put the aperture 40-50 pixels offset to upper, lower, left, and right of the source to extract four noise spectra. The final noise spectrum is the median of the four. We then fit the spectrum using either a single gaussian model or a double-peaked gaussian model with the python package \texttt{lmfit} and accept the better one (see Appendix \ref{sec:line_width} for details). Based on the fitting, we re-estimate the FWHM of the emission line (note as $W_{50}$\footnote{To clarify, the FWHM we used in \texttt{tclean} and $\lcotwo$ is a rough estimation from casa. While $W_{50}$ is a more carefully measured value. The differences between the two values are less than 10\% for all of our samples.}). Then the signal-to-noise ratio (S/N) is defined as the peak of the model divided by the standard deviation of the noise spectrum within $\pm2\times W_{50}$ of the emission line.

\section{results} \label{sec:results}
\subsection{CO measurements} \label{subsec:CO_measurement}
\begin{figure*}
\centering
\includegraphics[width=.8\linewidth]{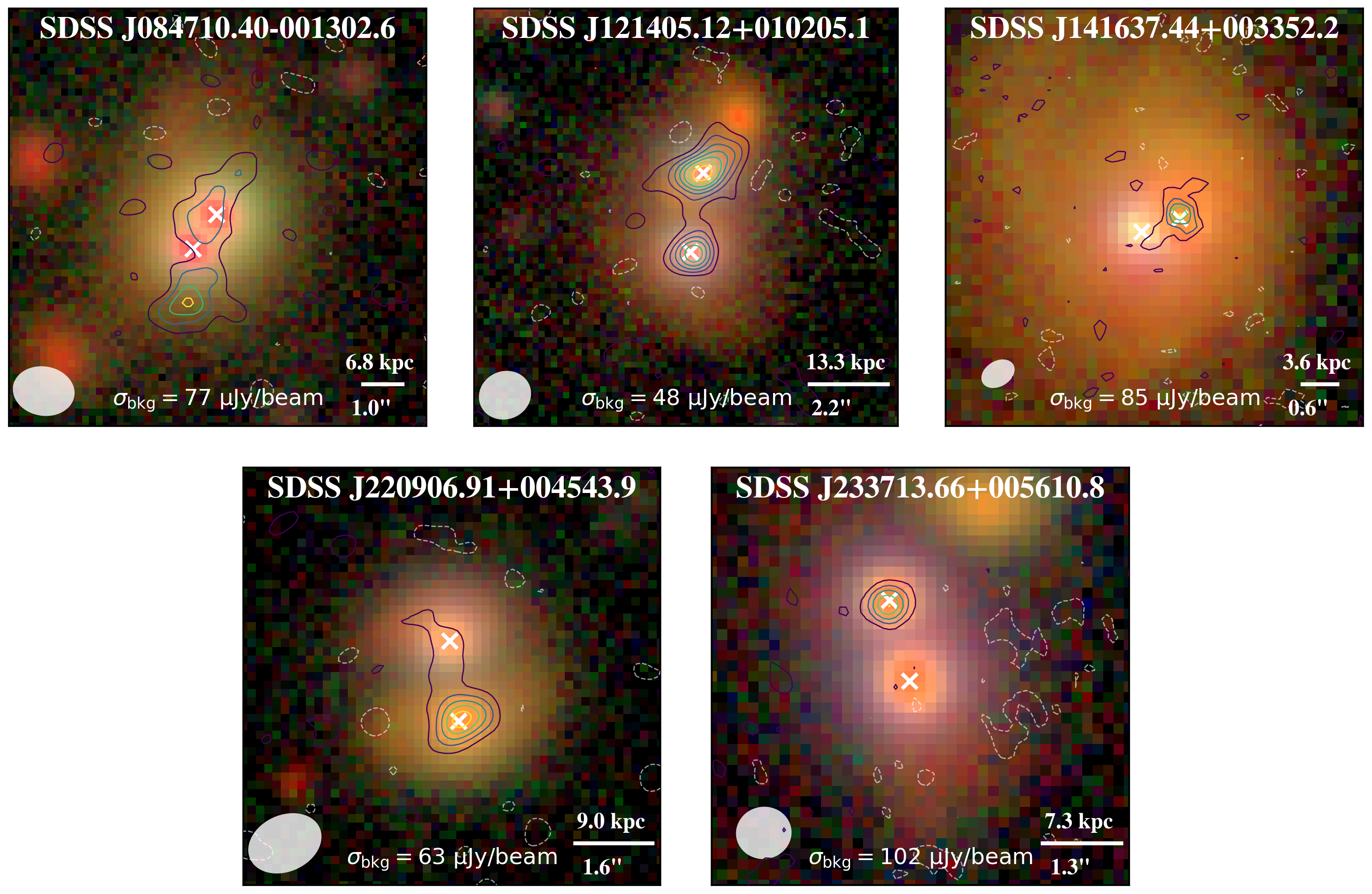}
\caption{HSC color ($gri$) images of five dual quasars with overlaid contours indicating the level of CO J=2--1 emission. The value of background noise ($\sigma_{\text{bkg}}$) is given at the bottom of each panel. The positive contours start at $2\sigma_{\text{bkg}}$ and increase in increments of $2\sigma_{\text{bkg}}$. The negative $2\sigma_{\text{bkg}}$ contours are plotted in white dashed lines. The optical centers of the point sources are pinpointed by the white crosses, and the scale bar at the bottom right indicates their projected separation. The beam size of ALMA is shown as the white ellipse at the bottom left.}
\label{fig:poststamps}
\end{figure*}

\begin{table*}
\caption{Summary of CO J=2-1 properties for observed dual quasars obtained through CASA processing. See Section \ref{sec:methods} for detailed methodology. Column (1): Abbreviated source names denoted by their relative positions within each pair. (2)-(3) Source positions estimated using \texttt{imfit} from the average CO line map. (4)-(5): Frequency center and $W_{50}$ (derived from equations \ref{eq:w50_double} and \ref{eq:w50_single}) of the CO J=2-1 line profile fitted with a single or double Gaussian (Section \ref{sec:line_width}). (6): Size of the objects in major axis FWHM and minor axis FWHM after deconvolution from the beam. (7): Integrated CO flux from the M0 map measured using \texttt{imfit}. For the non-detection sources, we assume them to be unresolved, and set the upper limits to $3\sigma_{\mathrm{bkg}}$ in one beam. (8): CO J=2-1 luminosity calculated from Equation \ref{eq:L_line}. (9): Molecular gas mass calculated assuming $R_{21}=0.62$ and $\alpha_{\mathrm{CO}}=3.1$ \citep{shangguan2020alma}. (10): Molecular gas-to-stellar mass ratio ($\umg=\mmg/M_*$), in which the stellar mass ($M_*$) is estimated from SED fitting (Table \ref{tab:optical}). For J0847-0013 and J1416+0033, sharing the same host galaxy, the total $M_*$ of the host is utilized as the denominator.}
\label{tab:measurements}
\begin{tabular}{ccccccccccc}
\hline
Name & RA & Dec & Freq. & $W_{50}$ & Size & $\scotwo$ & $\log~\lcotwo$ & $\log~\mmg$ & $\umg$\\
 & (hh:mm:ss) & (dd.mm.ss) & (GHz) & (km/s) & (kpc) & (mJy) & ($\mathrm{K\ km\ s^{-1}\ pc^2}$) & ($\mathrm{M_{\odot}}$) & \\
(1) & (2) & (3) & (4) & (5) & (6) & (7) & (8) & (9) & (10) \\
\hline
J0847N & 08:47:10.41 & -00:13:02.42 & 141.69 & $172\pm14$ & $18\pm4$ ; $4.6\pm2.2$ & $1.81\pm0.39$ & $9.49\pm0.09$ & $10.19\pm0.09$ & $19.6\pm4.2\%$\\
J0847S & 08:47:10.45 & -00:13:04.51 & 141.67 & $252\pm18$ & $14\pm4$ ; $5.3\pm1.8$ & $1.23\pm0.30$ & $9.51\pm0.11$ & $10.20\pm0.11$ & $20.2\pm4.9\%$\\
J1214N & 12:14:05.11 & 01:02:07.24 & 154.30 & $508\pm29$ & $7.9\pm1.0$ ; $3.0\pm0.7$ & $1.59\pm0.16$ & $9.75\pm0.04$ & $10.45\pm0.04$ & $84.5\pm8.5\%$\\
J1214S & 12:14:05.13 & 01:02:05.08 & 154.47 & $340\pm60$ & $3.7\pm1.0$ ; $2.6\pm1.3$ & $1.37\pm0.14$ & $9.49\pm0.04$ & $10.19\pm0.04$ & $34.8\pm3.6\%$\\
J1416W & 14:16:37.42 & 00:33:52.48 & 160.69 & $273\pm13$ & $2.9\pm0.6$ ; $1.9\pm0.5$ & $2.47\pm0.38$ & $9.46\pm0.07$ & $10.16\pm0.07$ & $20.7\pm3.2\%$\\
J1416E & -- & -- & -- & -- & -- & $<0.26$ & $<8.41$ & $<9.11$ & $<2.2\%$\\
J2209N & 22:09:06.92 & 00:45:43.47 & 159.34 & $260\pm40$ & $7.7\pm3.3$ ; $0.7\pm3.3$ & $0.58\pm0.23$ & $8.91\pm0.17$ & $9.61\pm0.17$ & $34.6\pm13.7\%$\\
J2209S & 22:09:06.90 & 00:45:42.26 & 159.35 & $340\pm40$ & $2.3\pm1.3$ ; $1.0\pm1.0$ & $0.79\pm0.12$ & $9.13\pm0.07$ & $9.83\pm0.07$ & $18.2\pm2.8\%$\\
J2337N & 23:37:13.70 & 00:56:12.00 & 134.95 & $128\pm8$ & $2.1\pm0.9$ ; $0.9\pm0.9$ & $1.25\pm0.16$ & $9.34\pm0.06$ & $10.04\pm0.06$ & $97.3\pm12.4\%$\\
J2337S & -- & -- & -- & -- & -- & $<0.31$ & $<8.73$ & $<9.43$ & $<3.1\%$\\
\hline
\end{tabular}
\end{table*}

Figure~\ref{fig:poststamps} shows the HSC color ($gri$) images \citep{lupton2004preparing} of the five dual quasars with the ALMA CO J=2--1 contours overlaid. The positive contours begin at $2\sigma_{\text{bkg}}$ and increase by a factor of 2, where $\sigma_{\text{bkg}}$ is the standard deviation of the residual map. The negative $2\sigma_{\text{bkg}}$ contours are plotted with the dashed lines. The white crosses mark the optical centers of the point sources, estimated from the $i$-band image of HSC, with the projected separations labelled at bottom right. Among the ten quasars, seven are associated with at least a 4$\sigma_{\text{bkg}}$ contour. From the 1D spectra, we estimate S/N of eight sources above $5\sigma$ \footnote{Note that this $\sigma$ is estimated from the 1D noise spectrum (Section \ref{subsec:observation}), while $\sigma_{\text{bkg}}$ is estimated from the residual map.}. We further highlight that the CO center of J0847S is displaced by $\sim$ 1.2\arcsec (8.2 kpc) from the optical center, while the other seven detections are co-spatial with the optical quasars. For the two non-detections, we assume them to be unresolved, and assign upper limits of $\scotwo$ to be $3\sigma_{\text{bkg}}$ in one beam (Table \ref{tab:measurements} column 7). Also from the extracted 1D spectral profile (detailed for each pair in Section \ref{sec:individuals}), we measure the frequency center and FWHM ($W_{50}$) of the CO emission line (Table \ref{tab:measurements} columns 4 and 5). The physical sizes of the identified sources are determined using \texttt{imfit} and reported in column (6). 
\par
\begin{figure*}
\centering
\includegraphics[width=.95\linewidth]{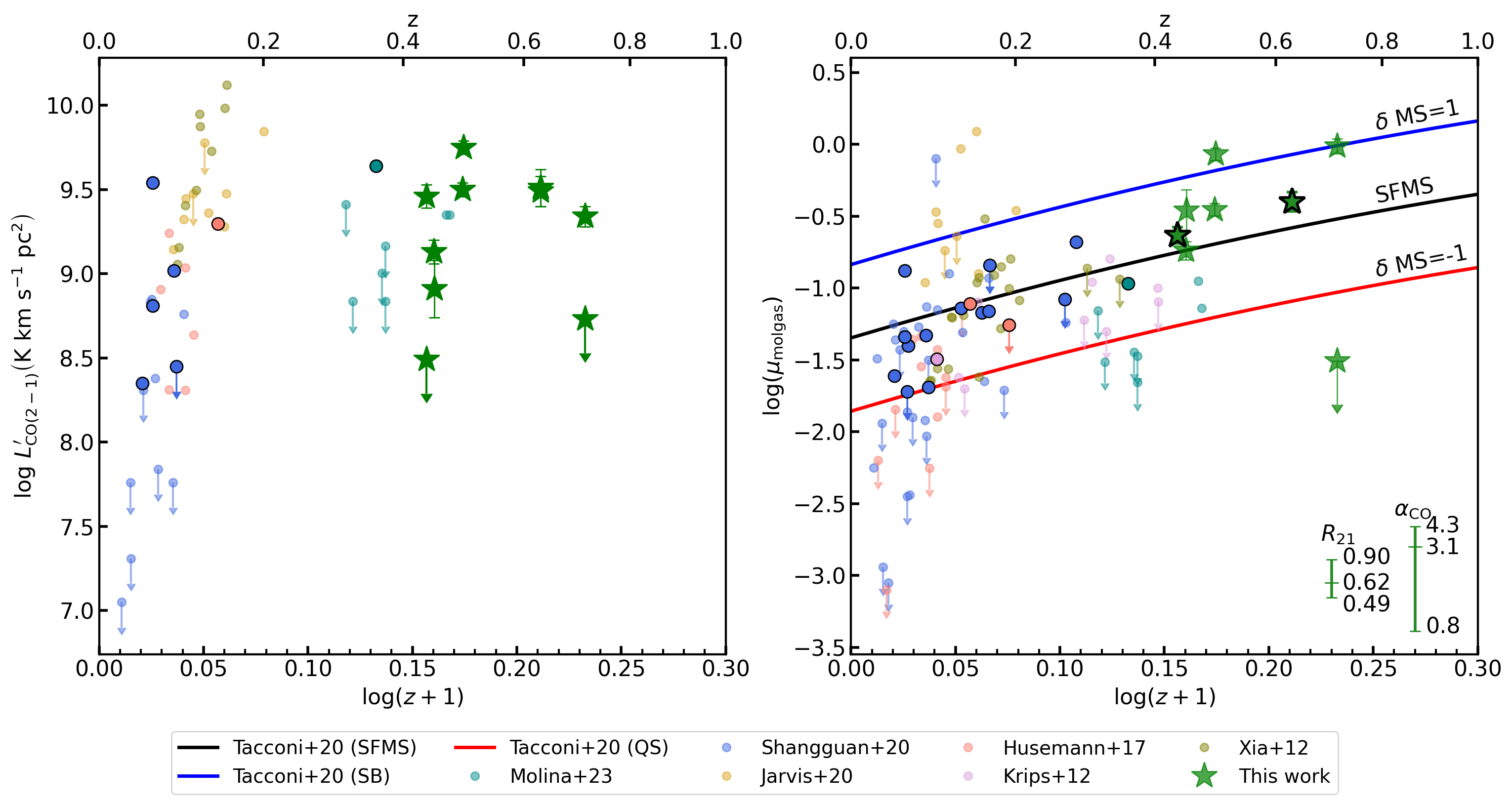}
\caption{Left: $\lcotwo$ as a function of redshift for the dual quasars in this study (green stars) and single quasars from the literature (colored dots). Right: Molecular gas to stellar mass ratio ($\umg = \mmg / M_*$) as a function of the redshift, using symbols similar to those in the left panel. Additionally, the figure includes the best-fitting results for the star-forming main sequence (SFMS), starburst (SB), and quiescent (QS) galaxies represented by black, blue, and red solid lines, respectively. The uncertainties associated with $R_{21}$ and $\alpha_{\mathrm{CO}}$ values are displayed at the bottom right. The pair of SDSS J0847-0013 and J1416+0033 are shown as a single data point (the green stars with black edges) because their host galaxies cannot be separated in HSC image, and we only have one $M_*$ (thus one $\umg$) measurement for each pair.}
\label{fig:gas_frac}
\end{figure*}

The CO luminosity $\lcotwo$ is calculated from $\scotwo$ using equation \ref{eq:L_line}, as listed in column (8). It is then plotted as a function of redshift in Figure \ref{fig:gas_frac} left panel, together with the single quasars in the literature. Based on a \texttt{logrank test}, we find no clear difference between the two populations  (see Section \ref{subsec:quench} for details).

\subsection{Gas masses and fractions}
We convert $\lcotwo$ to the CO J=1--0 luminosity ($\lcoone$) with the excitation correction factor $R_{21}=\lcotwo/\lcoone=0.62$, adopted from the average results of low-z PG quasars \citep{shangguan2020alma}. Among all of their sources with both $\lcotwo$ and $\lcoone$ detections, the highest $R_{21}$ value is 0.90, and the lowest is 0.49 (but also see \cite{carilli2013cool} that suggests $R_{21}=0.99$). $\lcoone$ is empirically related to $\mathrm{H_2}$ gas with a conversion factor $\alpha_{\mathrm{CO}}$ in a unit of $\mathrm{M_{\odot}}\left(\mathrm{K} \cdot \mathrm{km} \cdot \mathrm{s}^{-1} \cdot \mathrm{pc}^2\right)^{-1}$. We note that the value of $\alpha_{\mathrm{CO}}$ can vary with a dependence on gas density, temperature, and metallicity \citep[see][ for a comprehensive review]{bolatto2013co}. Various $\alpha_{\mathrm{CO}}$ values have been used in the literature, roughly ranges between $\alpha_{\mathrm{CO}}=0.8-4.3$ \citep[e.g.,][]{downes1998rotating,bolatto2013co}. For our dual quasars, we adopt $\alpha_{\mathrm{CO}}=3.1$, again following \cite{shangguan2020alma}. Based on our chosen values of $R_{21}$ and $\alpha_{\mathrm{CO}}$, we calculated $\mmg$ as listed in Table \ref{tab:measurements} column (9). Subsequently, we estimate $\umg = \mmg / M_*$ in Column (10) and show it as a function of redshift in the right panel of Figure \ref{fig:gas_frac}, in comparison with single quasars (colored dots) and inactive galaxies (solid curves) in the literature. The stellar mass ($M_*$) of our dual quasars is measured from SED fitting based on the host magnitudes after removing the point sources from the images (see Appendix \ref{sec:op_measurement} for a brief example). The uncertainties of $\umg$ by various selections of $\alpha_{\mathrm{CO}}$ and $R_{21}$ values are shown as the errorbars at bottom right.
\par
For the pair of SDSS J0847-0013 and J1416+0033, their host galaxies already merged into one. Therefore, we only have one measurement of $M_*$ for both. Their total $\umg$ are shown in Figure \ref{fig:gas_frac} right panel as the two green stars with black edges. Considering the total amount of molecular gas in all five systems, they all exceed $10^{10}~\mathrm{M_{\odot}}$, with $\umg$ falling between 10-60\% (Table \ref{tab:likelihood}). Overall, the total $\umg$ of dual quasars are similar or slightly elevated as compared to the star formation main sequence (SFMS) galaxies at the same redshift (black curve in Figure \ref{fig:gas_frac} right panel). The uncertainties and more detailed comparison with other types of sources are discussed in Section \ref{subsec:quench}.

\section{Details of individual pairs} \label{sec:individuals}
Here, we delve deeper into the specifics of each quasar pair. The results are encapsulated in one figure for each source. With Figure \ref{fig:0847_CO_map} as an example, our depiction follows a consistent structure: 

\begin{itemize}
\item Subaru/HSC color image with CO J=2--1 M0 map as contours overlaid (panel A).
\item Velocity (M1; panel b) and dispersion (M2; panel c) maps derived with \texttt{immoment} in units of km/s.
\item CO M0 image with flux levels as indicated (panel d). 
\item Best model fit (using \texttt{imfit}) based on the M0 data (panel e). 
\item Residual map by subtracting the model from the data (panel f). 
\item 1D spectra for both sources (panels g and h) extracted from the white dashed ellipses (apertures) in panel (d) shown as the grey histograms. The noise spectrum is plotted in cyan. The best fit result and 3$\sigma$ confidence region are shown as the red curve with shadows. $W_{50}$ of the emission line is marked as the blue vertical dashed lines. Based on optical spectroscopy, we estimated the optical redshift of the sources (see example in Appendix \ref{sec:op_measurement}). The expected frequency of CO J=2--1 line based on optical redshift is plotted as the cyan arrow above the 1D spectra. The S/N of the emission line is noted at the upper right side.
\end{itemize}

\subsection{SDSS J0847-0013} \label{subsec:J0847}
\begin{figure*}
\centering
\includegraphics[width=.95\linewidth]{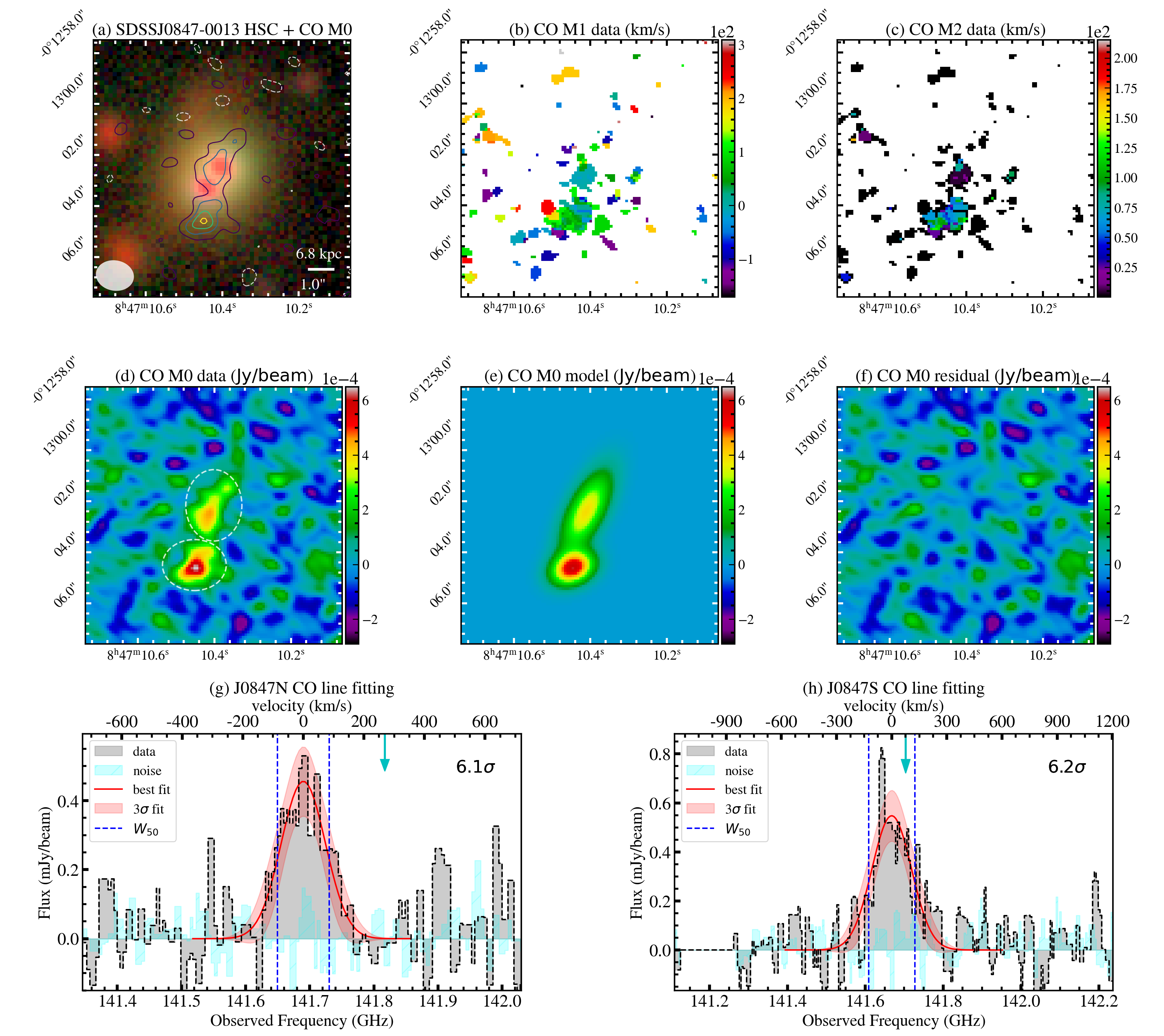}
\caption{CO (J=2--1) properties of SDSS J0847-0013. (a): CO M0 (intensity) map as described in Section \ref{subsec:observation}. The underlying color image is from Subaru/HSC $g$, $r$, and $i$ band data. The white ellipse at the bottom left indicates the beam size of our ALMA observation, and the projected optical separation between the two nuclei is shown as the scale bar at the bottom right with angular and physical scales. (b): CO M1 (velocity) map generated with \texttt{immoment} in unit of km/s. The velocity center is selected to be the center of the Gaussian model of the CO profile in panel (g). (c): CO M2 (dispersion) map generated with \texttt{immoment} in unit of km/s. (d): Same CO M0 map as in panel (a) with the flux level shown by the scale bar. (e): 2D gaussian model of CO M0 map generated from \texttt{imfit}. (f): CO residuals of the M0 map after subtracting the model in panel (e) from the data in Panel (d). (g) and (h): Emission line profiles of the two sources extracted from the regions marked in panel (d) together with the noise spectra extracted from offset regions from the sources, the best-fit model with 3$\sigma$ uncertainties from \textsc{lmfit}, and $W_{50}$ of the emission line. The arrows at the top indicate the expected position of the CO line based on the optical spectroscopic redshift.}
\label{fig:0847_CO_map}
\end{figure*}

First discovered by \cite{inada2008sloan} during their search for lensed quasars with Keck/LRIS, this system features two point sources separated by $1.0\arcsec$ at $z=0.626$ ($R_{\perp}=6.8$ kpc). The distinct profiles of the broad $\text{Mg \sc{ii}}$ emission lines in the two sources strongly indicate the nature of this system as a dual quasar rather than a lensed quasar. Subsequent investigations by \cite{silverman2020dual} determined black hole masses ($M_{\rm BH}$) of $10^{8.7}~\mathrm{M_{\odot}}$ for J0847-0013S and $10^{9.0}~\mathrm{M_{\odot}}$ for J0847-0013N based on $\text{Mg \sc{ii}}$ measurements utilizing Keck/LRIS. The host galaxy of this system is positioned between the two sources, and has an effective radius ($R_e$) of 0.94\arcsec\ (4.5 kpc) in the HSC $i$ band, as derived from the image decomposition results \citep{silverman2020dual}. The stellar mass is estimated to be $M_*\sim 10^{10.90}~\mathrm{M_{\odot}}$ through five-band spectral energy distribution (SED) fitting using the Code Investigating GALaxy Emission (\textsc{CIGALE}, \cite{boquien2019cigale}) tool (see Appendix \ref{sec:op_measurement} for a brief summary).
\par
For the results presented in Figure~\ref{fig:0847_CO_map}, the task \texttt{tclean} was centered on J0847-0013N at 141.69 GHz with FWHM=165 km/s, yielding a $6.1\sigma$ line detection, $\mmg=10^{10.19}~\mathrm{M_{\odot}}$, and $\umg=19.6\%$. The M1 and M2 map are generated between 141.54 GHz and 141.77 GHz, with stacked pixels above 1.2 mJy/beam (Figure~\ref{fig:0847_CO_map}bc). For J0847-0013S, \texttt{tclean} was centered at 141.67 GHz with FWHM=250 km/s, resulting in a $6.2\sigma$ line detection, $\mmg=10^{10.20}~\mathrm{M_{\odot}}$, and $\umg=20.2\%$. Both sources exhibit comparable molecular gas masses, and the extracted profiles for both can be adequately fitted with a single Gaussian with $W_{50}=172\pm14$ km/s for J0847-0013N and $W_{50}=252\pm18$ km/s for J0847-0013S (Figure~\ref{fig:0847_CO_map}gh).
\par
An intriguing aspect of this system is the offset of the CO center of J0847-0013S from its optical center. This offset is 1.2\arcsec\ in angular separation and 8.2 kpc in projected physical distance. There is no optical counterpart in the HSC image to this offset gas blob with 5$\sigma$ depth of $\sim$26.5 mag. Its position was not covered by the LRIS slit \citep{silverman2020dual}. The optical emission lines of J0847-0013N and J0847-0013S are offset by 240 km/s, while the CO line centers are only separated by approximately 42 km/s. The offset between the CO and optical redshifts is 277 km/s for J0847-0013N, and 55 km/s for J0847-0013S (as labelled by the cyan arrows in Figure~\ref{fig:0847_CO_map}gh, similar for the rest of the targets). The M1 and M2 maps of both sources are hardly resolved. We suggest two possible scenarios for this offset gas blob: either stripped by ram pressure or ejected by AGN feedback and (or) multi-body interaction (see Section \ref{subsec:evolution} for discussion).

\subsection{SDSS J1214+0102} \label{subsec:J1214}
\begin{figure*}
\centering
\includegraphics[width=.95\linewidth]{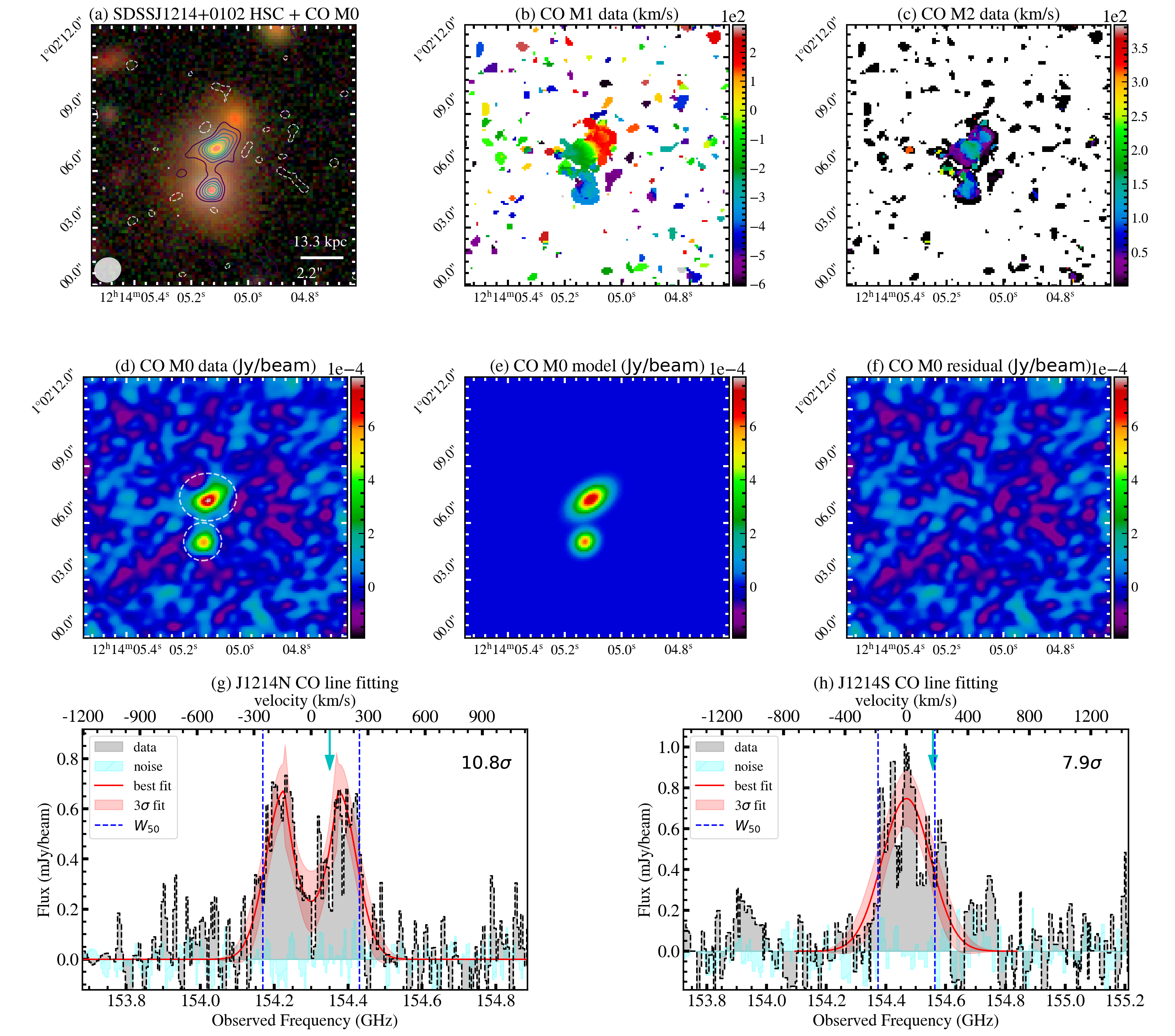}
\caption{CO properties of SDSS J1214+0102. The format is similar to Figure~\ref{fig:0847_CO_map}}
\label{fig:1214_CO_map}
\end{figure*}

This system, identified as a dual quasar at $z=0.493$, was confirmed using Keck/LRIS and has a separation of 2.2\arcsec\ ($R_{\perp}=13.3$ kpc) between the two quasars \citep{silverman2020dual}. Optical spectra revealed broad emission lines $\text{Mg \sc{ii}}$ and H$\beta$ in both sources. Based on H$\beta$, the BH masses are estimated to be $10^{9.1}~\mathrm{M_{\odot}}$ for J1214+0102N and $10^{8.7}~\mathrm{M_{\odot}}$ for J1214+0102S. The image decomposition resolved a distinct host galaxy for each quasar \citep{silverman2020dual} with $M_*=10^{10.52}~\mathrm{M_{\odot}}$ for J1214+0102N and $10^{10.65}~\mathrm{M_{\odot}}$ for J1214+0102S.
\par
For the CO results shown in Figure~\ref{fig:1214_CO_map}, \texttt{tclean} was centered at 154.29 GHz with FWHM=555 km/s for J1214+0102N. Its emission line is detected with 10.8$\sigma$. The CO luminosity translates to $\mmg=10^{10.45}~\mathrm{M_{\odot}}$, and $\umg=84.5\%$. This represents the largest molecular gas reservoir among the sources observed in this work. The M1 and M2 maps are generated between 154.14 GHz and 154.60 GHz with stacked pixels above 1.4 mJy/beam (Figure~\ref{fig:1214_CO_map}bc). Additionally, its 1D CO line shows a double-peaked Gaussian profile with $W_{50}=508\pm29$ km/s (Figure~\ref{fig:1214_CO_map}g). For J1214+0102S, \texttt{tclean} was centered at 154.47 GHz with FWHM=360 km/s. This reveals a $7.9\sigma$ line detection, $\mmg=10^{10.19}~\mathrm{M_{\odot}}$, and $\umg=34.8\%$. Its CO profile appears to be a single Gaussian with $W_{50}=340\pm60$ km/s (Figure~\ref{fig:1214_CO_map}h), with its center blueshifted from J1214+0102N by 330 km/s. The CO and optical redshifts are offset by 120 km/s for J1214+0102N, and 180 km/s for J1214+0102S.
\par
J1214+0102N stands out as the only source in this study exhibiting a distinct velocity gradient in its M1 map, as illustrated in Figure~\ref{fig:1214_CO_map}b. This structure is resolved by $\sim$ 2 beam sizes. The velocity zero point is set to be the center of the line profile of J1214+0102N at 154.30 GHz. The blueshifted component and the redshifted component are spatially offset by roughly one beam size in both the cubic data and the integrated M1 map. The $v/\sigma$ ratio of J1214+0102N is between 0.5 and 2. We consider this structure indicates either a merger front or a rotation disk in J1214+0102N (see discussion in Section \ref{sec:J1214N_disk}). 

\subsection{SDSS J1416+0033} \label{subsec:J1416}
\begin{figure*}
\centering
\includegraphics[width=.95\linewidth]{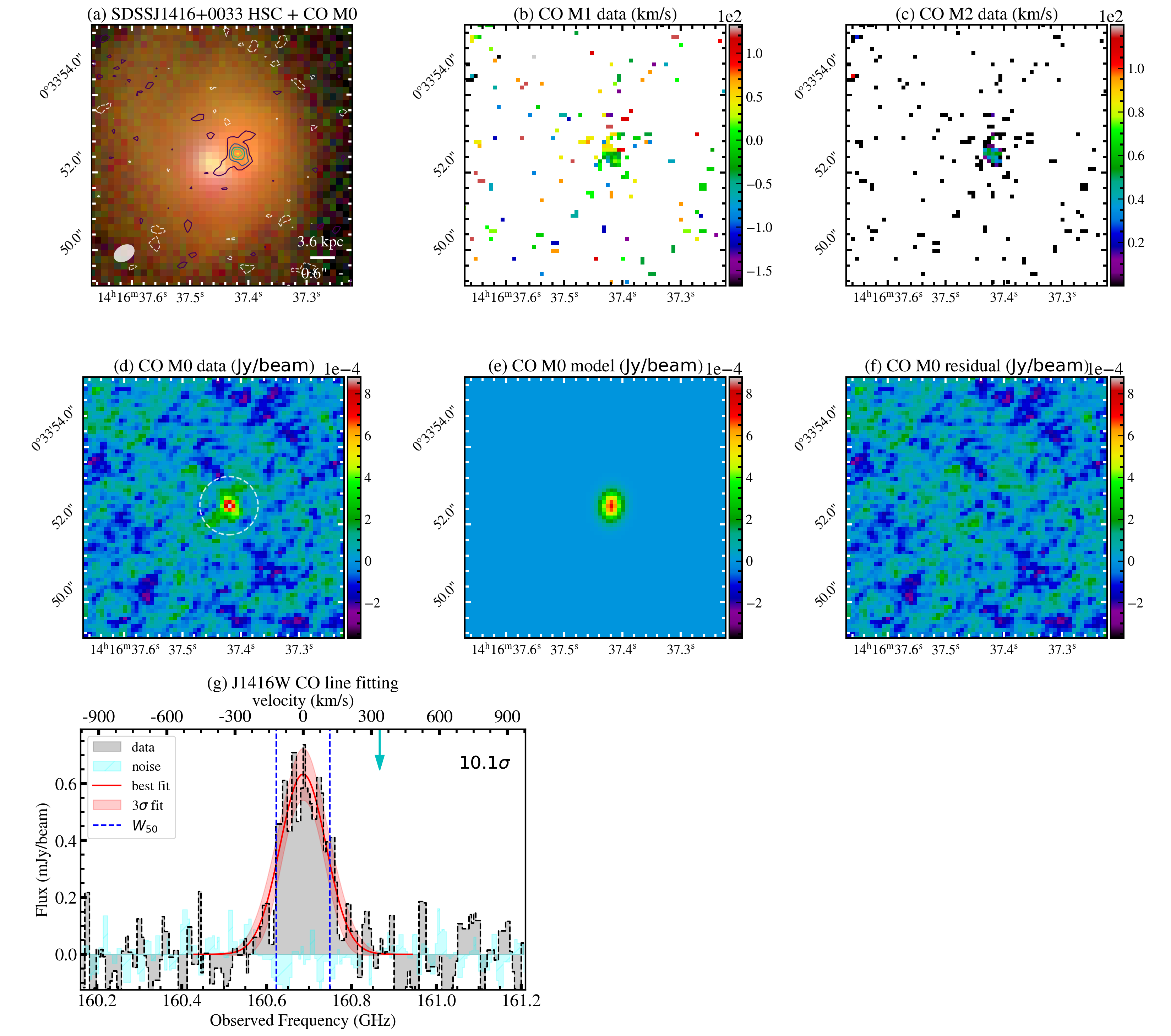}
\caption{CO properties of SDSS J1416+0033. The format is similar to Figure~\ref{fig:0847_CO_map}}
\label{fig:1416_CO_map}
\end{figure*}

This system, previously identified as a dual quasar at z = 0.434 in \cite{silverman2020dual}, is separated by 0.65\arcsec\ ($R_{\perp}$=3.9 kpc). Their spectra in Keck/LRIS were heavily blended; thus extraction from the extended wings of the 2D profile was performed for cleaner spectra. Consequently, broad $\text{Mg \sc{ii}}$ and H$\beta$ lines were prominent in J1416+0033E (left source in Figure~\ref{fig:1416_CO_map}a), whereas weaker in J1416+0033W (the right source). Subsequent Gemini/NIFS z-band IFU observation resolved the two nuclei, confirming the trend of broad H$\alpha$ emission being stronger in J1416+0033E and weaker in J1416+0033W. On the other hand, the existence of high-ionization narrow lines like [$\text{O \sc{iii}}$] and [$\text{Ne \sc{v}}$] \citep{yuan2016spectroscopic} in its spectrum suggested quasar-origin emission from J1416+0033W. Together with its red color in the HSC image, \cite{silverman2020dual} classified J1416+0033W as a type 1.5 quasar with some level of obscuration. However, the FWHM of the broad H$\alpha$, H$\beta$ and $\text{Mg \sc{ii}}$ lines of J1416+0033W is approximately double that of its type 1 companion, J1416+0033E, establishing it as the most massive BH in our sample ($M_{\rm BH} \sim 10^{9.1}\mathrm{M_{\odot}}$ with H$\alpha$). In comparison, the BH mass of J1416+0033E is $10^{8.6}\mathrm{M_{\odot}}$. Our decomposition analysis indicates that a single host galaxy was centered at the position of the two nuclei, with a stellar mass of $10^{10.84}\mathrm{M_{\odot}}$.
\par
For the CO results shown in Figure~\ref{fig:1416_CO_map}, \texttt{tclean} was centered on J1416+0033W (the type 1.5 companion) at 160.69 GHz with FWHM of 240 km/s, revealing a $10.1\sigma$ line detection with $W_{50} = 237 \pm 13$ km/s (Figure~\ref{fig:1416_CO_map}g). The CO and optical redshifts are offset by 335 km/s, which is the largest offset among our samples. The CO measurements correspond to $\mmg = 10^{10.16}~\mathrm{M_{\odot}}$, and $\umg=20.7\%$. The M1 and M2 maps are generated between 160.62 GHz and 160.78 GHz with stacked pixels above 1.7 mJy/beam (Figure~\ref{fig:1416_CO_map}bc). On the other hand, J1416+0033E (the type 1 companion) is undetected, we estimate its $3\sigma$ upper limit of $\scotwo<0.26$ mJy, which corresponds to $\mmg < 10^{9.11}~\mathrm{M_{\odot}}$, and $\umg<2.2\%$. 
\par
The size of this source observed by ALMA is close to the beam size, with a major axis measuring less than one-third of the \sersic\ radius derived from the HSC $i$ band image (2.9 kpc compared to 9.2 kpc). Both the M1 and M2 maps of this source appear unresolved. From this, we infer that the molecular gas in this system is relatively compact compared to the stellar components, predominantly concentrated in J1416+0033W. This CO emitting gas may also be the cause of the obscuration of this companion.

\subsection{SDSS J2209+0045} \label{subsec:J2209}
\begin{figure*}
\centering
\includegraphics[width=.95\linewidth]{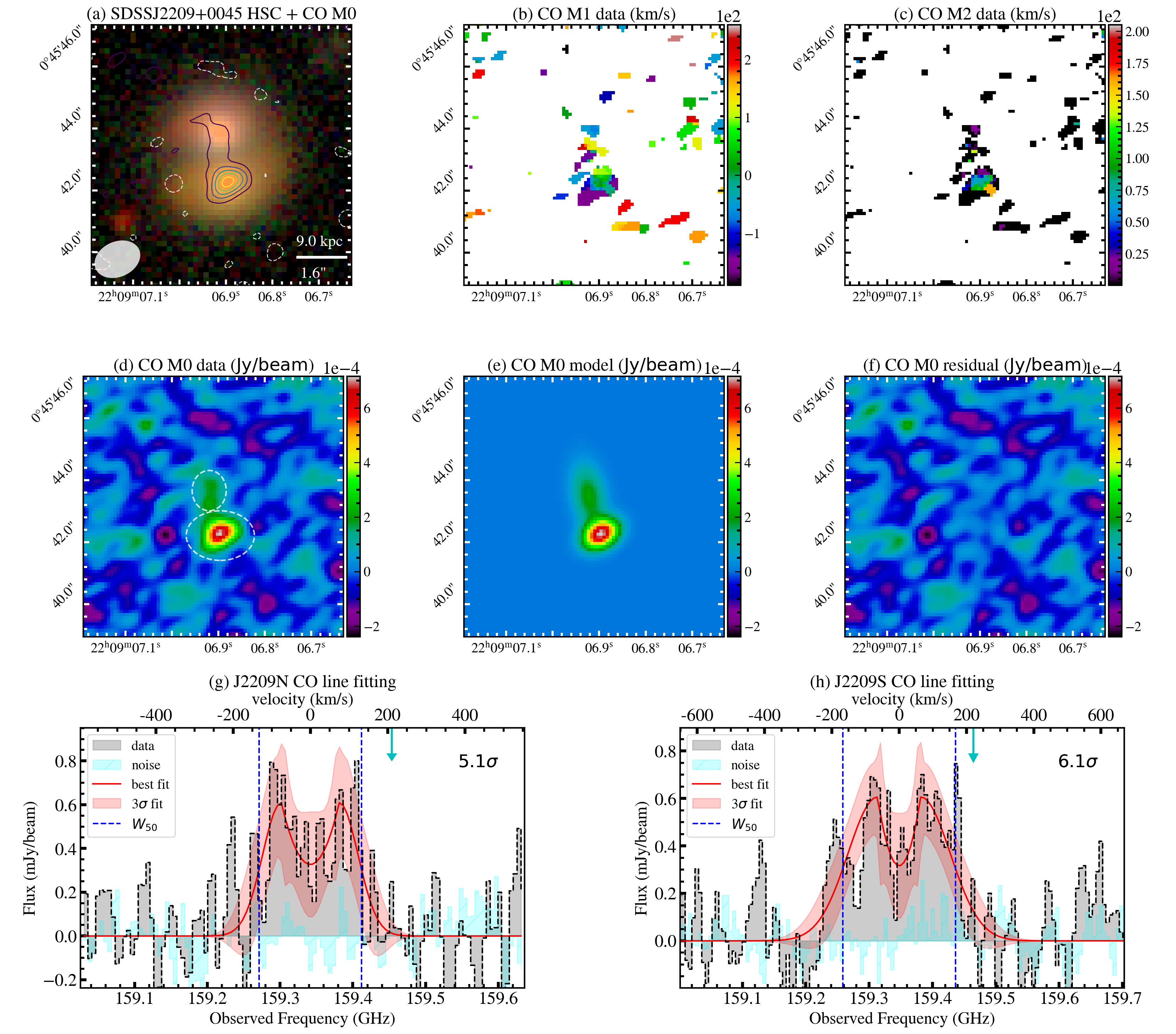}
\caption{CO properties for SDSS J2209+0045. The format is similar to Figure~\ref{fig:0847_CO_map}.}
\label{fig:2209_CO_map}
\end{figure*}

This system was identified in \cite{tang2021optical} as a dual quasar at $z=0.446$ with a separation of 1.63\arcsec\ ($R_{\perp}$=9.2 kpc). Subsequent spectroscopy with Subaru/FOCAS revealed the presence of broad H$\alpha$ and H$\beta$ lines in both sources. From the H$\alpha$ observations, we estimate $M_{\rm BH}$ as $10^{8.0}\mathrm{M_{\odot}}$ for J2209+0045N and $10^{7.5}\mathrm{M_{\odot}}$ for J2209+0045S. The host galaxies of the two sources are distinguishable in the HSC image with $M_* = 10^{10.07}\mathrm{M_{\odot}}$ for J2209+0045N and $10^{10.57}\mathrm{M_{\odot}}$ for J2209+0045S.
\par
For the CO results presented in Figure~\ref{fig:2209_CO_map}, we centered \texttt{tclean} on J2209+0045S at 159.35 GHz with FWHM=330 km/s. A 2D double Gaussian model fitting in \texttt{imfit} estimated $\scotwo=0.79$ mJy for J2209+0045S, corresponding to $\mmg = 10^{9.83}~\mathrm{M_{\odot}}$ and $\umg=18.2\%$ (Figure~\ref{fig:2209_CO_map}e). For J2209+0045N, we centered \texttt{tclean} at 159.24 GHz with FWHM=270 km/s, which reveals $\scotwo=0.58$ mJy, $\mmg = 10^{9.61}\mathrm{M_{\odot}}$, and $\umg=34.6\%$. 
\par
The extracted spectra for both sources exhibit double-peaked features (Figure~\ref{fig:2209_CO_map}gh). Based on a double-peaked gaussian model, the line is detected with 5.1$\sigma$ and $W_{50}=260\pm40$ km/s for J2209+0045N, and 6.1$\sigma$ and $W_{50}=340\pm40$ km/s for J2209+0045S. The optical emission lines are offset from the CO line center by 207 km/s for J2209+0045N, and 228 km/s for J2209+0045S. The M1 and M2 maps were generated between 159.22 GHz and 159.46 GHz, with stacked pixels above 1.5 mJy/beam (Figure~\ref{fig:2209_CO_map}bc). However, both of the sources are unresolved under the large beam size.
\par
An intriguing feature of this system is the $2\sigma$ gas bridge between the two companions (Figure~\ref{fig:2209_CO_map}a). A similar structure also exists in J0847-0013 (Figure~\ref{fig:0847_CO_map}a) and J1214+0102 (Figure~\ref{fig:1214_CO_map}a). These emissions could be originated from the overlapped halo of the two galaxies (e.g., see the case of PACS-787 in \cite{silverman2018concurrent}, but also see \cite{tan2024fitting} that interpreter this as an artifact of Fourier transform from the $uv$-plane). 

\subsection{SDSS J2337+0056} \label{subsec:J2337}
\begin{figure*}
\centering
\includegraphics[width=.95\linewidth]{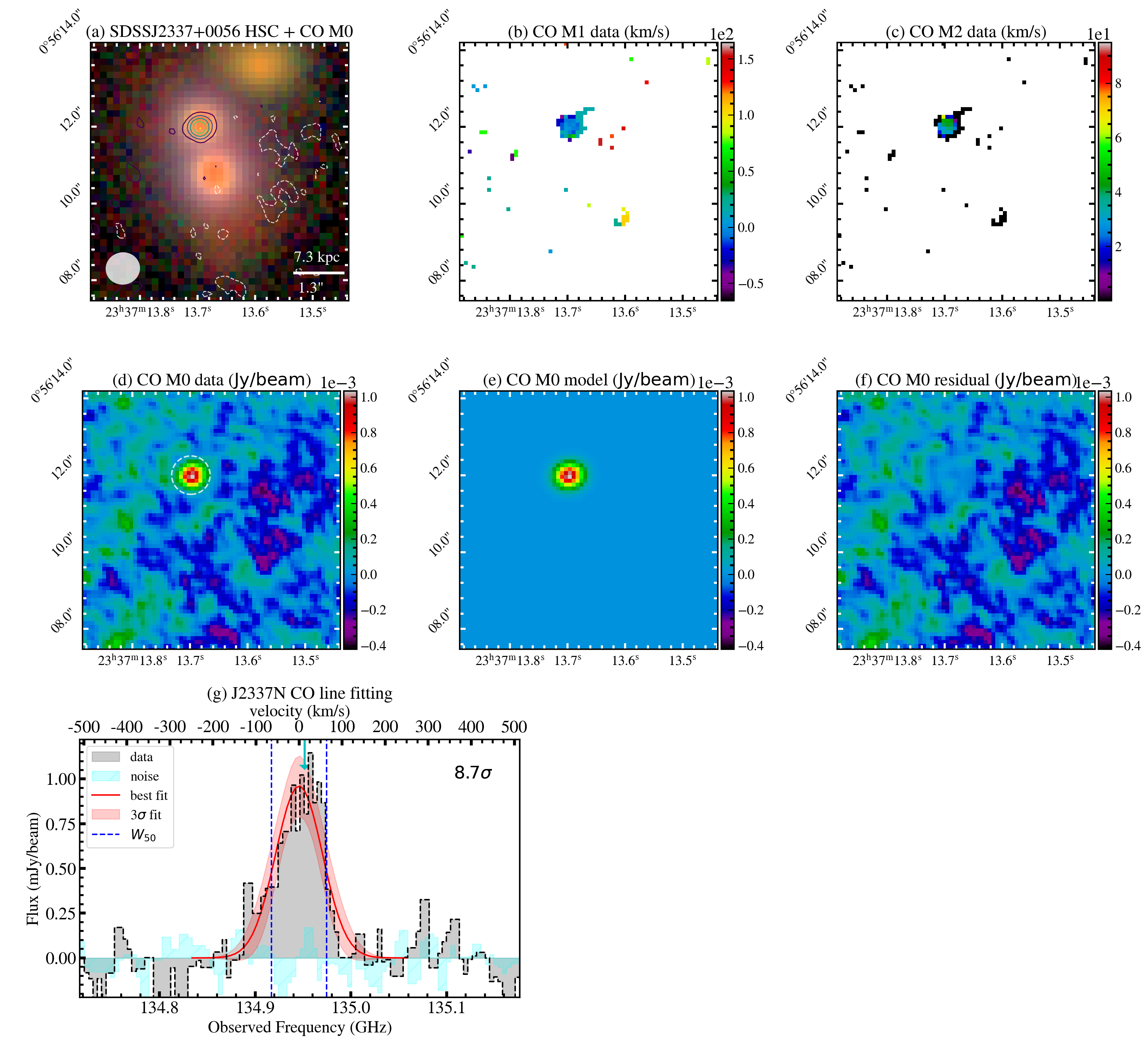}
\caption{CO properties for SDSS J2337+0056. The format is similar to Figure~\ref{fig:0847_CO_map}.}
\label{fig:2337_CO_map}
\end{figure*}
This system was first reported in \cite{tang2021optical} as a dual quasar at z=0.708 separated by 1.34\arcsec\ ($R_{\perp}$=7.6 kpc). Follow-up spectroscopy with Gemini/GMOS revealed broad H$\beta$ lines in both sources, from which we estimate $M_{\rm BH}$ as $10^{7.9}\mathrm{M_{\odot}}$ for J2337+0056N and $10^{8.2}\mathrm{M_{\odot}}$ for J2337+0056S. Also, broad and blue-shifted [OIII] components are observed in both sources, indicating an outflow velocity of 490 km/s for J2337+0056N and 260 km/s for J2337+0056S \citep{tang2021optical}. The host galaxy weighs $10^{10.05}\mathrm{M_{\odot}}$ for J2337+0056N and $10^{10.94}\mathrm{M_{\odot}}$ for J2337+0056S (Table \ref{tab:optical}).
\par
For the CO results shown in Figure~\ref{fig:2337_CO_map}, we centered \texttt{tclean} on J2337+0056N at 134.95 GHz with FWHM=130 km/s. This yields a $8.7\sigma$ detection of the line with $W_{50}=128\pm8$ km/s. The optical and CO redshifts are almost identical, with an offset of 18 km/s (Figure~\ref{fig:2337_CO_map}g). The line luminosity translates to $\mmg = 10^{10.04}\mathrm{M_{\odot}}$ and $\umg=97.3\%$. The M1 and M2 maps are generated between 134.87 GHz and 134.98 GHz with stacked pixels above 1.5 mJy/beam (Figure~\ref{fig:2337_CO_map}bc).
\par
Similar to J1416+0033, the CO emission of this system is compact and concentrated on one source of the pair, not the primary SDSS quasar. We estimate the 3$\sigma_{\text{bkg}}$ upper limit of J2337+0056S to be $\scotwo<0.31$ mJy, which corresponds to $\mmg < 10^{9.4}\mathrm{M_{\odot}}$ and $\umg<3.1\%$. Although both nuclei appear unobscured in optical spectroscopy, the molecular gas distribution turns out to be very asymmetric in this system. One possibility is that, the gas-rich quasar might be much younger than the gas poor quasar, which has depleted its surrounding gas during previous evolutionary stages.

\section{Discussion} \label{sec:discussion}
Our results reveal that dual quasars typically possess substantial molecular gas reservoirs, with $\mmg$ between $10^{9.6-10.5}~\mathrm{M_{\odot}}$ and $\umg$ between $18-97\%$. In this section, we first analyze these findings statistically to qualitatively determine whether these systems are quenched. Next, we will explore the varied distribution of molecular gas among our dual quasars. This analysis aims to offer insights into the behavior of quasars during galaxy mergers and to relate these observations to the broader context of BH-host coevolution.

\subsection{Do dual quasars show signs of gas depletion?} \label{subsec:quench}
We aim to determine whether dual quasars show evidence of gas depletion, which may hint at ongoing quenching of subsequent star formation. To address this question, we look into the CO J=2--1 luminosity ($\lcotwo$) and the molecular gas to stellar mass ratio ($\umg$) as shown in Figure~\ref{fig:gas_frac}. The comparison samples include single PG quasars at $z<0.1$ \citep{shangguan2020alma} and $z\lesssim0.5$ \citep{molina2023lack}, bright ($L_{\rm bol}>10^{44}\mathrm{ergs^{-1}}$) Hamburg/ESO quasars \citep{husemann2017integral}, infrared ultraluminous quasars \citep{xia2012molecular}, and type-2 quasars \citep{krips2012co,jarvis2020high}.
\par
We first compare the $\lcotwo$ of single and dual quasars. The \cite{molina2023lack} sample provides the closest comparison to our dual quasars in terms of redshift and black hole properties ($L_{\mathrm{bol}}$ and $M_{\mathrm{BH}}$). However, a key difference is the instrument used in their study (NOEMA). \cite{molina2023lack} reported weak correlations between the molecular gas properties and the black hole properties of their samples, suggesting that AGN feedback is not a significant factor in quenching. We applied the \texttt{logrank test} (Mantel–Cox test, \cite{mantel1966evaluation}) using the \texttt{lifelines} Python package to compare the $\lcotwo$ values of their samples with those of our dual quasars. Accounting for the upper limits as well, the test yielded a p-value of 0.50, indicating that these two populations are indistinguishable in terms of $\lcotwo$. We refrain from showing the correlations between the CO properties and the black hole properties of our dual quasars, as they may not be at the same stage of the merger process.
\par
The same comparison samples are also plotted in the right panel of Figure~\ref{fig:gas_frac}. For the $\umg$ estimations, we adopted their assumptions for $\alpha_{\mathrm{CO}}$ and $R_{21}$ values, which are more appropriate for their specific types of objects. \cite{krips2012co} and \cite{xia2012molecular} lack stellar mass measurements for individual sources. We assumed $M_{\text{total}} = M_{\text{molgas}} + M_* = 10^{11} \mathrm{\mathrm{M_{\odot}}}$ based on the statements in their respective studies.
\par
In addition to the quasars, we have incorporated measurements from inactive galaxies into this figure. Specifically, the solid curves represent the best-fitting results of \cite{tacconi2020evolution}. The fitting encompasses gas mass detections of 2,052 SFGs from the literature, spanning redshifts from 0 to 5.2, stellar masses between $10^{9-12.2}~\mathrm{M_{\odot}}$, and star formation rates (SFR) between $10^{-1.5-3.75}~\mathrm{M_{\odot}}$/yr. Their fitting function to the molecular gas-to-stellar mass ratio is a function of redshift, relative specific star formation rate (sSFR), stellar mass, and effective radius:

\begin{align}
\log \left(\umg\right) &= A + B \times [\log (1+z)-F]^2 \nonumber \\
&\quad + C \times \log \left[\mathrm{sSFR} / \mathrm{sSFR}\left(\mathrm{MS}, z, M_*\right)\right] \nonumber \\
&\quad + D \times [\log \left(M_*\right)-10.7]
\label{eq:gas_frac}
\end{align}

We apply their best-fit results $A=0.06$, $B=-3.33$, $F=0.65$, $C=0.51$, $D=-0.41$ to the function. In the right panel of Figure~\ref{fig:gas_frac}, we present the reproduced scaling relation of $\umg=\mmg / M_*$ with redshift for the star formation main sequence (SFMS) galaxies as the black solid curve. Let $\delta \text{MS}=\log \left[\mathrm{sSFR} / \mathrm{sSFR}\left(\mathrm{MS}, z, M_*\right)\right]$. The black curve corresponds to $\umg(z=0-1,\delta \text{MS}=0,\log M_*=10.7)$, where $\log~M_*/\mathrm{M_{\odot}}=10.7$ is the median value of local galaxies \citep{saintonge2011cold}, and the sSFR of SFMS galaxies is defined in \cite{speagle2014highly}. We vary $\delta \text{MS}$ to 1 and -1, representing the blue curve for starburst (SB) galaxies and the red curve for quiescent (QS) galaxies. 
\par
Continuing with our analysis, we assess the likelihood of our dual quasars being positioned above the blue curve (SB-like), between the blue and red curves (SFMS-like), or below the red curve (QS-like). For each dual quasar, we calculate the total $\mmg$ and $M_*$ by summing the values of the two companions. This allows us to determine the total molecular-to-stellar mass ratio for each pair ($\mu^{\text{dual}}_{\text{molgas}}$). SDSS J0847-0013 and J1416+0033 are already presented in this manner and are shown as the two green stars with black edges in the right panel of Figure~\ref{fig:gas_frac}. The other three pairs are shown separately, their total $\mu^{\text{dual}}_{\text{molgas}}$ will be in between the two components. We consider uncertainties in measuring $\umg$, which include: (1) CIGALE measurement uncertainties of $M_*$, (2) CASA measurement uncertainties of $\mmg$, and (3) systematic uncertainties of $\mmg$ related to $R_{21}$ and $\alpha_{\text{CO}}$. We assume the systematic uncertainties to be $R_{21} = 0.62^{+0.28}_{-0.13}$ and $\alpha_{\text{CO}} = 3.1^{+1.2}_{-2.3}$, as represented by the green error bars at the bottom right of the right panel of Figure~\ref{fig:gas_frac}. Furthermore, we assume the probability density function (PDF) follows a split normal distribution.
\begin{equation}
\begin{array}{ll}
f\left(x ; \mu, \sigma_{\text{low}}, \sigma_{\text{up}}\right)=A \exp \left(-\frac{(x-\mu)^2}{2 \sigma_{\text{low}}^2}\right) & \text { if } x<\mu \\
f\left(x ; \mu, \sigma_{\text{low}}, \sigma_{\text{up}}\right)=A \exp \left(-\frac{(x-\mu)^2}{2 \sigma_{\text{up}}^2}\right) & \text { otherwise }
\end{array}
\label{eq:mu_up_low}
\end{equation}
where $A=\sqrt{2 / \pi}\left(\sigma_{\text{low}}+\sigma_{\text{up}}\right)^{-1}$. Then the 95\% confidence interval can be converted to an asymmetric upper and lower $1\sigma$ range by dividing the 95\% bounds by a factor of two. This yields $R_{21}=0.62^{+0.14}_{-0.07}$ and $\alpha_{\text{CO}}=3.1^{+0.6}_{-1.2}$. Subsequently, these three uncertainties are combined together, yielding the upper and lower $1\sigma$ uncertainties for $\mu^{\text{dual}}_{\text{molgas}}$ for our dual quasars, as presented in Table \ref{tab:likelihood} column (4).
\par
For comparison, we calculate the corresponding $\umg^{\text{SFMS}}$ for each of our sources using equation \ref{eq:gas_frac}, i.e., $\umg^{\text{SFMS}}=\umg(z=z_{\text{source}}, ~\delta \text{MS}=0, ~\log M_*=10.7)$. The results are presented in Table \ref{tab:likelihood}, column (5), with the upper and lower limits representing $\delta \text{MS}=\pm1$, respectively. We classify values between the upper and lower limits as "SFMS-like", those above the upper limit as "SB-like", and those below the lower limit as "QS-like". Using Equation \ref{eq:mu_up_low}, we estimate the probabilities of $\umg^{\text{dual}}$ falling into these three regions. These probabilities, denoted as $P_{\mu}$, are listed in Table \ref{tab:likelihood}, column (6), in the order of SB-like, SFMS-like, and QS-like.
\par
As a result, $\umg^{\text{dual}}$ of our dual quasars are slightly higher than the $\umg^{\text{SFMS}}$ of SFMS galaxies, except J2337+0056. Considering both systematic and measurement uncertainties, all five pairs show at least 70\% probability of being either SFMS-like or SB-like. However, the $\umg$ upper limits of J1416+0033E and J2337+0056S (Table \ref{tab:measurements}, column 11) fall below the lower limit of $\umg^{\text{SFMS}}$ at their respective redshifts. Thus, at this merger phase, our dual quasars do not present significant evidence of cold molecular gas depletion when viewed as a combined sample. Nevertheless, a few individual quasars within dual systems exhibit little to no CO-emitting gas. 
\par
On the other hand, the separations between the nuclei can be approximated as a probe of the merger stage. Comparing the projected physical separations between the nuclei (Table \ref{tab:likelihood} column 2) with the total $\mmg^{\text{dual}}$ and $\umg^{\text{dual}}$ (Table \ref{tab:likelihood} column 3, 4). We find no clear correlations between the molecular gas properties the separations, although the most separated case J1214 is the gas-richest system. It is possible that the gas depletion time scale is longer than the dynamical time scale of merging \citep{capelo2015growth}, and most of the gas are depleted after the final coalescence \citep{hopkins2006unified}. Nevertheless, the samples in this work has limited statistics, and more follow-up observations on dual quasars with various separations will be favored to test the depletion scenario.

\begin{table*}
\caption{The projected physical separation ($R_{\perp}$), total molecular gas mass ($\log~\mmg^{\text{dual}}$) and total molecular-to-stellar mass ratio ($\umg^{\text{dual}}$) for each pair of the dual quasars. In comparison with the best-fit results of SFMS galaxies galaxies from \citet{tacconi2020evolution} (equation \ref{eq:gas_frac}) at the same redshift ($\umg^{\text{SFMS}}$). The lower and upper $1\sigma$ uncertainties for $\umg^{\text{dual}}$ account for both measurement and systematic errors (see text for details). That of $\mu^{\text{SFMS}}_{\text{molgas}}$ represents the variation of $\delta \text{MS}=\pm1$ in equation \ref{eq:gas_frac}. Column (6) provides the percentage probabilities that the $\umg^{\text{dual}}$ of our dual quasars fall above the upper limit (SB-like), between the lower and upper limit (SFMS-like), and below the lower limit (QS-like) of $\mu^{\text{SFMS}}_{\text{molgas}}$, assuming a split normal distribution PDF of the uncertainties (equation \ref{eq:mu_up_low}).}
\label{tab:likelihood}
\begin{tabular}{cccccc}
\hline
Name & $R_{\perp}$ & $\log~\mmg^{\text{dual}}$ & $\mu^{\text{dual}}_{\text{molgas}}$ & $\mu^{\text{SFMS}}_{\text{molgas}}$ & $P_{\mu}(\%)$\\ 
 & (kpc) & ($\mathrm{M_{\odot}}$) & (\%) & (\%) &  (SB, SFMS, QS)\\
(1) & (2) & (3) & (4) & (5) & (6)\\
\hline
\vspace{5pt}
J0847 & 6.8 & $10.50\pm0.07$ & $39.5^{+22.0}_{-26.8}$ & $26.2^{+58.6}_{-18.1}$ & 1.95, 85.94, 12.10\\
\vspace{5pt}
J1214 & 13.3 & $10.64\pm0.03$ & $56.1^{+18.4}_{-28.5}$ & $20.2^{+45.2}_{-14.0}$ & 30.75, 65.23, 4.02\\
\vspace{5pt}
J1416 & 3.6 & $10.20\pm0.06$ & $22.8^{+10.0}_{-13.3}$ & $17.7^{+39.6}_{-12.2}$ & 0.02, 90.27, 9.70\\
\vspace{5pt}
J2209 & 9.0 & $10.03\pm0.08$ & $22.2^{+8.2}_{-11.9}$ & $18.2^{+40.8}_{-12.6}$ & 0.00, 91.83, 8.17\\
\vspace{5pt}
J2337 & 7.3 & $10.14\pm0.05$ & $13.9^{+6.7}_{-8.6}$ & $30.2^{+67.6}_{-20.9}$ & 0.00, 70.16, 29.84\\
\hline
\end{tabular}
\end{table*}

\subsection{J1214+0102N: Merger or disk?} \label{sec:J1214N_disk}
In Section~\ref{subsec:J1214} and Figure~\ref{fig:1214_CO_map}, we highlighted the gradient feature observed in the velocity map. Here, we discuss potential explanations for this structure. Firstly, its $v/\sigma$ ratio is close to unity, supporting evidence of a merger (e.g., see the numerical analysis in \citealp{lapi2018dramatic}, and a case of local dual AGN \citealp{feruglio2013high}). The merger front likely aligns along the minor axis of the CO M1 map of J1214N (Figure~\ref{fig:1214_CO_map}b). Notably, the HSC image reveals another red galaxy to the northwest of J1214N, separated by 2.0\arcsec or 8.4 kpc (Figure~\ref{fig:1214_CO_map}a). Although we lack spectroscopic data for this galaxy, the image shows tidal features between it and J1214N. It is undetected with ALMA, suggesting it is a quiescent galaxy if at the same redshift. Therefore, this system may reside in a compact over-dense region within a 20 kpc scale.
\par
The biconical outflow driven by AGN feedback can also produce double-peaked emission lines and velocity gradient features in both molecular and ionized gas phases \citep{cicone2014massive,comerford2018origin}. However, such outflows typically result in asymmetrical line profiles due to the likely asymmetric geometry of the outflow bicone \citep{nevin2018origin}. In the case of SDSS J1214+0102N, the nearly identical peaks in the CO emission line (Figure~\ref{fig:1214_CO_map}g) do not align with the expected asymmetry of an outflow scenario. Similarly, galaxy mergers usually generate asymmetric line profiles because the brightness of the two galaxies involved is typically not identical \citep{maschmann2020double}.
\par

\begin{figure}
\centering
\includegraphics[width=.95\linewidth]{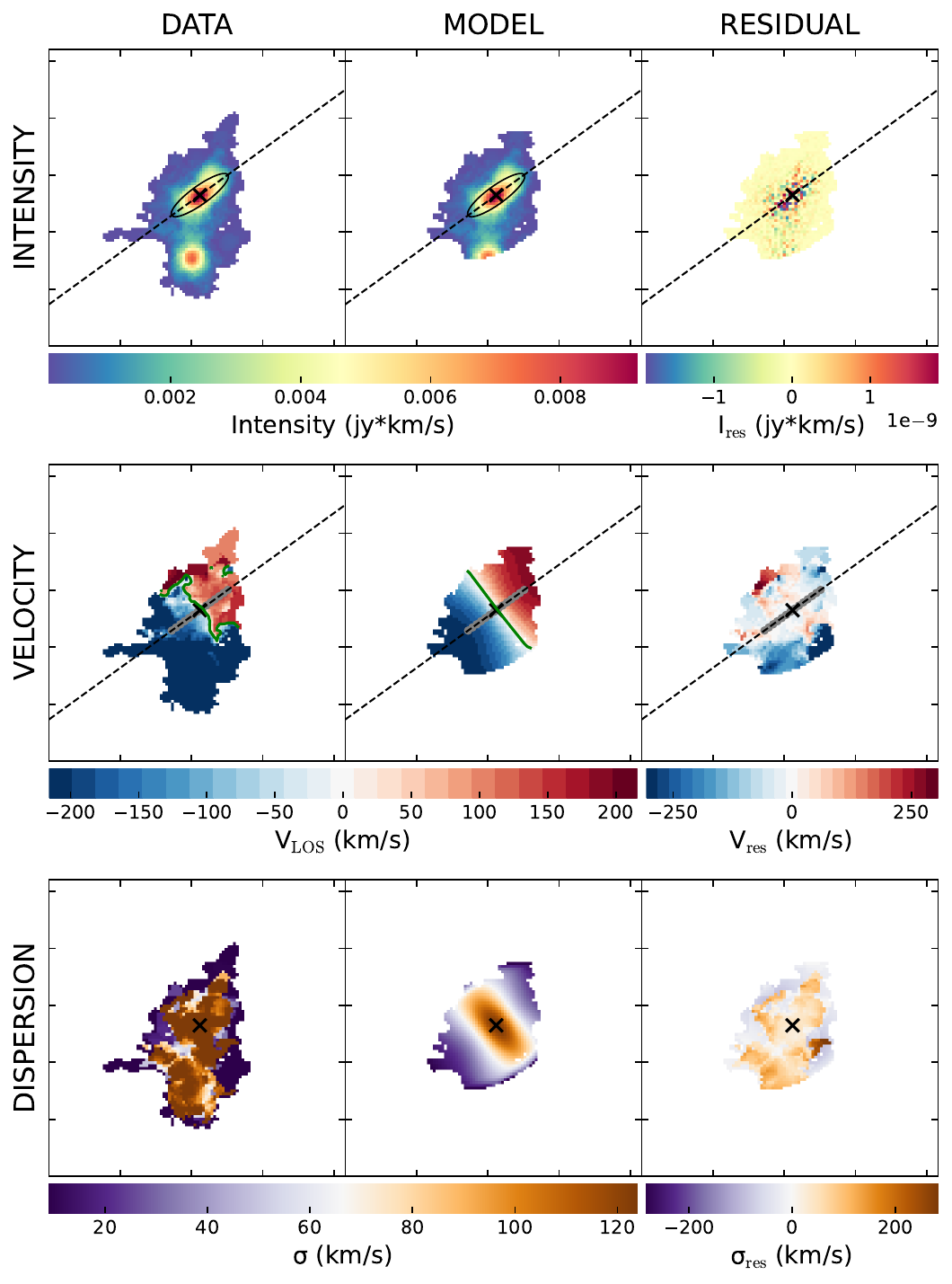}
\caption{$^{\rm{3D}}$Barolo output set for SDSS J1214+0102. The results for M0, M1, and M2 maps are from top to bottom. From left to right, the data, fitting model, and residuals are displayed. The black crosses mark the center position of the disc model. Only the northern sources are used for fitting, as circled by the black elliptical aperture. The dashed lines indicate the angle of the major axis.}
\label{fig:1214_barolo}
\end{figure}

Therefore, although disfavored by the $v/\sigma$ ratio, we still consider a rotating disk as an alternative scenario, which is known to produce symmetric double-peaked emission lines \citep{storchi2017double,maschmann2020double,maschmann2023origin}. To test this hypothesis, we perform a kinematic analysis of this feature using the tool named 3D-Based Analysis of Rotating Objects via Line Observations \citep[$^{\rm{3D}}$Barolo,][]{teodoro20153d}. In brief, $^{\rm{3D}}$Barolo uses a number of concentric rings to fit a model of a rotating disc based on emission line data cubes. This tool has been extensively used to study the gas kinematics of galaxies \citep[e.g., ][]{ginolfi2020alpine,maddox2021mightee}. However, careful tuning is required to produce a successful fit, especially for low S/N and low-resolution data. The essential point for this particular case is to remove J1214S from modeling, which otherwise will be considered as a part of the disc. The \texttt{SEARCH} task of $^{\rm{3D}}$Barolo is based on DUCHAMP \citep{whiting2012duchamp}, which produces a 3D source mask. We raise the \texttt{SNRCUT} of detection to 7, \texttt{GROWTHCUT} to 4, and \texttt{MINCHANNELS} to four to only keep J1214+0102N in the model and reduce noise pixels. We then use the \texttt{imfit} results on M0 map to further constrain the model, following a similar practice in \cite{banerji2021resolving}. We fix the model center (\texttt{XPOS} and \texttt{YPOS}) and position angle (\texttt{PA}) to the \texttt{imfit} results. Then decide the ring model parameters as follows:
\begin{equation}
\mathrm{NRADII} = \mathrm{round}(\frac{\mathrm{FWHM_{major}}}{\mathrm{FWHM_{minor}/2.5}})
\end{equation}
\begin{equation}
\mathrm{RADSEP} = \frac{\mathrm{FWHM_{major}}}{\mathrm{NRADII}}
\end{equation}
where \texttt{NRADII} is the number of rings, which is 7 in our case. \texttt{RADSEP} is the separation (width) of the rings, which is 0.187. These two parameters ensure the fitting is focused on the center region, neither over-resolving the data nor including too much surrounding noise. We leave rotation velocity (\texttt{VROT}), velocity dispersion (\texttt{VDISP}), inclination angle (\texttt{INC}), and Scale-height of the disc (\texttt{Z0}) to be free parameters after assigning initial guesses from the M1 and M2 maps ($\texttt{VROT}=200$, $\texttt{VDISP}=40$, $\texttt{INC}=71$, $\texttt{Z0}=0$). In particular, for the inclination angle, the initial value is calculated from the axial ratio of the galaxy following \cite{holmberg1958photographic}:
\begin{equation}
\mathrm{INC}=\cos ^{-1}\left(\sqrt{\frac{q^2-q_0^2}{1-q_0^2}}\right)
\end{equation}
where q is the ratio of the semi-minor to the semi-major axis of the galaxy (i.e., $\mathrm{FWHM_{minor}}/\mathrm{FWHM_{major}}$) achieved from \texttt{imfit}, q0 is assumed to be 0.2 \citep{pierce1988distances}. 
\par
The inclination angle finally converges to 75.4 degrees. As a cross check, we compared with the 2D decomposition results on Subaru/HSC $i$-band image using \textsc{GaLight} \citep{ding2022galight}, in which we find $q=0.31$, i.e., $\texttt{INC}=76$ degrees for J1214+0102N. The output model and residual from $^{\rm{3D}}$Barolo are shown in Figure~\ref{fig:1214_barolo}. From the top to the bottom are the intensity (M0) map, velocity (M1) map, and dispersion (M2) map, respectively. From left to right are corresponding observational data, fitting models, and the residuals. The best-fit results suggest an almost constant \texttt{VROT} $\sim$ 225 km/s from inner to outer rings. After subtracting the instrumental broadening ($\sigma_{\text {instr}}=W_{\text {ch}} / \sqrt{2\ln 2}$ as defined in $^{\rm{3D}}$Barolo, where the channel width $W_{\text {ch}}$ is 10km/s for our case), the velocity dispersion decreases roughly linearly from 32km/s for the innermost ring to 4km/s for the second outermost ring. 
\par
With \texttt{VROT} and the size of the major axis ($a_{\mathrm{major}}$, Table~\ref{tab:measurements} column 6), we calculate the dynamical mass of J1214+0102N with the following formula:
\begin{equation}
\left(\frac{M_{\mathrm{dyn}}}{\mathrm{M_{\odot}}}\right)=1.16 \times 10^5\left(\frac{\mathrm{VROT}}{\mathrm{km}~\mathrm{s}^{-1}}\right)^2\left(\frac{1.5\times a_{\mathrm{major}}}{\mathrm{kpc}}\right)
\end{equation}
The factor 1.5 is applied to cover the faint emissions of the disk \citep[see ][]{wang2013star, izumi2021subaru}. As a result, we obtained $M_{\mathrm{dyn}}\sim 10^{10.84}~\mathrm{M_{\odot}}$ for J1214+0102N. Assuming $M_{\mathrm{dyn}} = M_{\mathrm{gas}} + M_*$, taking the SED-measured stellar mass (Table \ref{tab:optical}, $M_* \sim 10^{10.52}\mathrm{M_{\odot}}$), this will leave $M_{\mathrm{gas}} \sim 10^{10.56} M_{\odot}$. For comparison, the molecular gas mass we measured from CO J=2-1 is $M_{\mathrm{molgas}} \sim 10^{10.45}M_{\odot}$, which thus takes $\sim 77.6\%$ of the total gas mass.
Overall, both the merger and the rotation disk scenarios are consistent with our current data. We are planning for better resolution and multi-wavelength observations to reveal the nature of this source.

\subsection{Insights on galaxy evolution} \label{subsec:evolution}
We aim to derive insights from our findings regarding the broader context of galaxy evolution with a focus on the connection between galaxy mergers and quasars. The conventional evolutionary framework posits a substantial role for gas-rich major mergers in driving quasar activity \citep{hopkins2008cosmological}. However, subsequent observational studies have generated much debate \citep[e.g.,][]{gabor2009active,cisternas2010bulk,silverman2011impact,boehm2013agn,villforth2014morphologies,villforth2017host,ellison2019definitive,marian2019major,marian2020significant,zhao2022relation}. The intricacies of this debate arise from various factors, largely the selection criteria for quasars such as the consideration of spectral coverage, luminosity, redshift and the diverse methods employed to define mergers (e.g., visual inspection, morphological parameters, machine learning).
\par
This highlights the advantages of studying dual quasars: They are well-defined mergers. Their merger stages can be approximated by the separations between the two optical nuclei, and the SMBH properties are available for both galaxies. This puts them in a well-constrained framework to discuss the co-evolution scenario. We first review the characteristics of our dual quasar sample. The bolometric luminosities, estimated from their monochromatic 5100 Å luminosities, span the range of $10^{44}$ to $10^{45.5}$ erg/s, positioning them in the middle of the faint end of the quasar luminosity function at $z\sim1$ \citep{shen2020bolometric}. The BH and host masses of these systems are similar to the parent SDSS DR14 quasar sample \citep{tang2021optical}. Regarding the representativity of this phase, we refer to the overall merger fraction for SDSS-HSC cross-matched quasars within this luminosity range at $0.2<z<0.8$ \citep{tang2023morphological}. According to the morphological approach in this work, the overall quasar merger fraction falls between 10\% and 25\%. With the Horizon-AGN simulation predicting a dual AGN fraction (among all AGN) with $L_{\text{bol}}>10^{44}$ erg/s, $R_\perp<30$ kpc, and $z\sim0.5$ to be around 2\% \citep{volonteri2022dual}, our dual quasars might represent a phase that accounts for 8\%-20\% of the lifetime of a quasar undergoing a merger event. 

\par
In Section \ref{subsec:quench}, we compared the global amount of molecular gas in each quasar pair with single quasars and inactive galaxies. We find no significant differences between these populations. Thus even there are two active quasars, their impact on the global gas environment seem to be insignificant. Next, we consider whether the molecular gas in each quasar has any preferential distribution. For example, does the molecular gas always associated with the brighter ($L_{\text{bol}}$), more massive ($M_{\text{BH}}$), or more actively accreting ($\lambda_{\text{Edd}}$) SMBH? In SDSS J1416+0033, J2209+0045, and J2337+0056, the CO detections are strongly biased towards one of the companions. Based on our measurements of their optical properties (Table \ref{tab:optical}), the detected J1416+0033W represents the brighter, more massive, and less actively accreting companion; J2209+0045S is the fainter, less massive, and less actively accreting companion; J2337+0056N is the fainter, less massive, and more actively accreting companion. The results, however, turn out to be random, displaying no discernible preference toward the SMBH properties. 
\par
It has been widely accepted that AGN feedback is a self-regulated process that maintains a balance between the growth of the SMBH and its surrounding environment \citep[see reviews by][]{fabian2012observational,morganti2017many}. Hence, the absence of correlations between the amount of molecular gas and transient properties such as $L_{\text{bol}}$ and $\lambda_{\text{Edd}}$ is not surprising. Meanwhile, as a time-integrated property, $M_{\text{BH}}$ is more likely to exhibit correlations with the total amount of molecular gas in quasars. This expectation also naturally arises from the combination of the $M_{\text{BH}}$-$M_*$ relation \citep{kormendy2013coevolution} and the $M_*$-SFR relation \citep{dave2008galaxy}. However, given the substantial uncertainties inherent in the virial methods used to measure $M_{\text{BH}}$ and conversion factors to molecular gas mass, claiming such a correlation necessitates a larger sample.
\par
We further ask whether the biased distribution of molecular gas could be the result of the merger? i.e., could one of the SMBHs accretes most of the molecular gas from the system via the interaction? To explore this hypothesis, we examine the evolutionary trajectories of similar systems in hydrodynamical simulation studies. \cite{capelo2015growth} investigated the evolution of physical properties of SMBHs and host galaxies during mergers, spanning a stochastic stage well before the galaxies interact to a remnant stage after the BHs coalesce below the stellar softening length. In particular, \cite{capelo2015growth} traced the gas mass within radii of 0.1, 1, and 10 kpc around both the primary and secondary BHs in galaxy mergers. For a BH mass ratio of 1:4 (similar to SDSS J1416+0033), their simulation show that the global gas content (within 1 or 10 kpc radius, close to the scales that we are pointing) remained almost constant for both BHs throughout the entire merger process. Similar results were obtained in their simulation of a 1:2 merger (similar to SDSS J2209+0045 and J2337+0056), where the global gas amounts were scarcely affected during the merger process. Consequently, we interpret the presence or absence of gas detection in our dual quasars as a reflection of the intrinsic properties of each galaxy prior to their current state. The ``gas bridge" structures, as observed in SDSS J2209+0045, play a role in transferring minor amount of gas that fuels a gas-poor quasar. Nonetheless, it is crucial to reiterate that our detection limit for $\mmg$ is somewhat above $10^9~\mathrm{M_{\odot}}$, which is the threshold defining ``gas-poor" and ``non-detection" in this work.
\par
In addition to transferring matter, another effect of galaxy mergers is suggested to be compaction of the galaxy \citep[][]{barnes1996transformations,zolotov2015compaction,tacchella2016evolution,chabanier2020formation}. In recent observations, \cite{molina2021compact} measured the CO J=2--1 half-light radius ($R_{1/2}$) of six $z\leq0.06$ PG quasars using ALMA and found a median value of 1.8 kpc. This is systematically smaller than inactive SFGs with similar global molecular gas properties from the EDGE-CALIFA survey \citep{bolatto2017edge} which are typically above 3 kpc. The sizes of our dual quasars are given in Table \ref{tab:measurements} column (6). Assuming Gaussian models, $R_{1/2}$ is half of the FWHM \citep[see other options in][]{anders2006accurate}. As a result, for J1214+0102S, J1416+0033W, J2209+0045S, and J2337+0056N, $R_{1/2}$ is less than 2 kpc thus appearing to be more compact than inactive SFGs. However, without a statistically significant number of well-matched control samples of non-merger and post-merger inactive SFGs and quasars, it remains unclear whether AGN feedback or mergers contribute to the compaction. Additionally, we should note that size measurements of merging galaxies could be affected by their tidal structures.
\par
As an alternative to a compaction scenario, SDSS J0847-0013 shows an extended molecular gas distribution as compared to the stellar emission. Its southern gas clump has an offset from the optical center by 8.2 kpc. This is uncommon in our sample and is rarely reported in the literature. Yet, it is a case from which we can infer the gas dynamics of the merger. We consider two scenarios for this extended and offset emission: (1) stripped by ram pressure (or in-falling) and (2) an ejected gas blob by multi-body interaction or AGN feedback. 
\par
First, ram pressure takes the form of $P_r \approx \rho_e v^2$, where $\rho_e$ is the external medium (the medium of another galaxy, in the case of a merger) density, and $v$ is the velocity of the galaxy \citep{gunn1972infall}. Although the SMBHs and stellar components are also pushed by the same $P_r$, their cross-sections are much smaller than the gas components, thus the resisting force is weaker. This leads to the retarded gas component. In N-body and hydrodynamic simulations, \cite{kapferer2008influence} found that up to $\sim$ 50\% of the total gas in the merger can be stripped by ram pressure and retarded tens of kpc behind the stellar discs of the galaxies after a fairly long time ($\sim$ 500 Myr) of interaction. Observational evidence of such an offset exists in, for example, the merging galaxy cluster Abell 754 \citep{zabludoff1995collision}, and Abell 2146 \citep{canning2012riding}. As the gas gets compressed by the ram pressure, new stars will be formed inside. Parts of the gas will also fall back onto the host galaxy. In both ways of ram pressure stripping and in-falling gas blob, the velocity direction should be towards the center.
\par
In contrast, if the gas blob is moving away from the center, then it is likely being ejected from this system by multi-body interaction or AGN feedback. A similar case was reported in \cite{carniani2017agn}. In one of their $z=2.4$ quasars, they reported blueshifted CO J=3--2 emission centered 1.3 kpc away from the quasar center, which overlaps with the broad $\mathrm{[O_{III}]\lambda 5007}$ emission stacked from VLT/SINFONI data cube. They suggest it to be evidence of an outflow triggered by AGN feedback. However, our case in J0847-0013S is much more extreme. The spatial offset is beyond the effective size of the host galaxy (8.2 kpc vs. 6.8 kpc, measured from HSC $i$-band). And the amount of offset molecular gas is comparable to the gas that remains at the center ($10^{10.20}~\mathrm{M_{\odot}}$ vs. $10^{10.19}~\mathrm{M_{\odot}}$). It is doubtful whether AGN feedback would be strong enough to cause such a violent ejection, although the Keck/LRIS spectrum of this source does show slightly asymmetric $\mathrm{[O_{III}]\lambda 5007}$ to the blue side \citep{silverman2020dual}.
\par

\section{Conclusions} \label{sec:conclusion}
In this work, we present the first results of using ALMA to observe cold molecular gas (CO J=2--1) from five closely-separated dual quasars with $R_{\perp}<20$ kpc, $L_{\rm bol}\gtrsim10^{44}~\mathrm{erg~s}^{-1}$ at $0.4<z<0.8$. Our main findings are summarized as follows:
\begin{itemize}
    \item Eight of ten quasars in five dual systems exhibit CO J=2-1 line detections above $5\sigma$, with $\scotwo>0.5$ mJy. These detections collectively suggest a typical molecular gas mass ($\mmg$) between $10^{9.6-10.5}~\mathrm{M_{\odot}}$, and molecular gas-to-stellar mass ratios ($\umg$) spanning $18-97\%$, based on assumptions of $R_{21}=0.62$ and $\alpha_{\rm CO}=3.1$ (Figure \ref{fig:poststamps}, Table \ref{tab:measurements}). The two non-detection cases have 3$\sigma$ upper limits of $\scotwo \sim 0.3$ mJy, which correspond to $\mmg < 10^{9.5}~\mathrm{M_{\odot}}$ and $\umg<5\%$. 
    \item The distribution of molecular gas in our dual quasars exhibits diverse characteristics: J0847-0013 has an offset gas blob located 8.2 kpc from the optical center; J1214+0102N is the only source showing a clear velocity gradient, which could indicate either a merger or a rotating disk; J1416+0033 is the only type1-type1.5 pair in our sample, with CO gas detected only for the obscured companion. In contrast, J2337+0056, a type1-type1 pair, also shows CO gas detected in only one companion; J2209+0045 shows a $2\sigma$ gas bridge, potentially indicating matter transfer between the two companions (Section \ref{sec:individuals}).
    \item A logrank test comparing the CO luminosities ($\lcotwo$) of our dual quasars with single quasars at similar redshifts reveal no statistically significant differences (Figure \ref{fig:gas_frac}). Considering each pair as one system, their total molecular-to-stellar mass ratios ($\mu^{\text{dual}}_{\text{molgas}}$) all exceed 10\%. Taking in both measurement and systematic uncertainties of $\mu^{\text{dual}}_{\text{molgas}}$, all of these five pairs have $>$70\% probability to be consistent with the star-forming galaxy population at the same redshift (Figure \ref{fig:gas_frac}, Table \ref{tab:likelihood}).
\end{itemize}
In summary, the molecular gas environments of dual quasars are rich and diverse, even within a narrow parameter space and uniform selection. The two main questions that remain are: (1) At which stage of galaxy mergers will the molecular gas be depleted? (2) What are the underlying physical mechanisms driving the diversity in molecular gas distribution? To address these questions, a larger sample across different redshifts and with various separations needs to be observed using multi-wavelength and high-resolution techniques. Combining a broader range of observational data with simulations will provide deeper insights into these phenomena.

\section*{Acknowledgements}
This paper makes use of the following ALMA data: ADS/JAO.ALMA\#2021.1.01233.S ALMA is a partnership of ESO (representing its member states), NSF (USA) and NINS (Japan), together with NRC (Canada), MOST and ASIAA (Taiwan), and KASI (Republic of Korea), in cooperation with the Republic of Chile. The Joint ALMA Observatory is operated by ESO, AUI/NRAO and NAOJ.
\par
ST and MB acknowledge funding from the Royal Society via a University Research Fellowship to MB and associated Research Fellows Enhancement Awards. CB gratefully acknowledges support from the Forrest Research Foundation. Special thanks go to the East Asian ALMA Regional Center (EA ARC) and the UK ALMA Regional Centre (UK ARC) nodes for their responsive support on the data achievement and processing of this work

\section*{Data Availability}
The raw data is public on ALMA archive with project ID: 2021.1.01233.S. The reprocessed data are available in the online supplementary material of this paper.



\bibliographystyle{mnras}
\bibliography{main_text} 




\appendix

\section{Line width measurements} \label{sec:line_width}
As discussed in Section \ref{subsec:observation}, we utilized the spectral profile tool within \texttt{casaviewer} to estimate the Full Width at Half Maximum (FWHM) of the emission line for \texttt{tclean}. However, this tool employs only simple Gaussian models, which may not always accurately represent the true profile of the CO emission line. Here, we provide a detailed strategy of our approach for fitting the CO J=2-1 line in our sophisticated measurements of the $W_{50}$ values in Table \ref{tab:measurements}.
\par
\cite{tiley2016tully} conducted a comprehensive assessment of various fitting functions to determine their effectiveness in reproducing the FWHM of simulated galaxy spectra across a broad parameter space, such as the amplitude-to-noise ratio (A/N), inclination, and rotation velocity. Their findings indicate that a parabolic function flanked by two mirrored half-Gaussians, known as a ``symmetric Gaussian Double Peak function", exhibited the smallest bias across most of the parameter space. This function is expressed as follows:
\begin{equation}
\begin{aligned}
f(v)= \begin{cases}A_{\mathrm{G}} \times \mathrm{e}^{\frac{-\left[x-\left(c-w\right)\right]^2}{2 \sigma^2}} & x<c-w, \\ 
A_{\mathrm{C}}+a\left(x-c\right)^2 & c-w \leq x \leq c+w, \\ 
A_{\mathrm{G}} \times \mathrm{e}^{\frac{-\left[x-\left(c+w\right)\right]^2}{2 \sigma^2}} & x>c+w\end{cases}
\label{eq:double_peak}
\end{aligned}
\end{equation}
where x is the observed frequency, c is the line center, $w>0$ is the half-width of the parabola, $\sigma>0$ is the width of the two half-Gaussians centered at $c\pm w$, they share the same peak flux $A_G>0$. Besides, we set a constraint of $A_{\mathrm{C}}\leq 0.8A_{\mathrm{G}}$ to keep the fitting stable.
\par
However, not all galaxies exhibit intrinsically double-peaked line profiles, as this characteristic relies on factors such as inclination and the gas distribution \citep[e.g.,][]{lavezzi1997recovering, davis2011atlas3d}. Consequently, we initially fitted both a double-peaked Gaussian model and a standard single Gaussian model to the extracted CO J=2-1 lines of our sources. The fitting process utilized the Python package \texttt{lmfit}, employing the Levenberg-Marquardt (leastsq) algorithm for optimization. To assess the quality of the fitting, we established three criteria: the Akaike information criterion \citep[AIC,][]{akaike1998information}, Bayesian information criterion \citep[BIC,][]{schwarz1978estimating}, and reduced chi-square. The preferred model was determined based on having lower values in at least two of these three criteria.
\par
Based on the accepted model, we estimate the $W_{50}$ of the CO lines as follows. For the double-peaked Gaussian model:
\begin{equation}
W_{50}=2(w+\sqrt{2 \ln 2} \sigma)
\label{eq:w50_double}
\end{equation}
and for the standard single Gaussian model:
\begin{equation}
W_{50}=2\sqrt{2 \ln 2} \sigma
\label{eq:w50_single}
\end{equation}
We first perform the above fitting in frequency space to get the precise position of the CO line, then convert the width $W_{50}$ to velocity, as reported in Table \ref{tab:measurements} column (5).

\section{Optical measurements} \label{sec:op_measurement}
\begin{figure*}
\centering
\includegraphics[width=.7\linewidth]{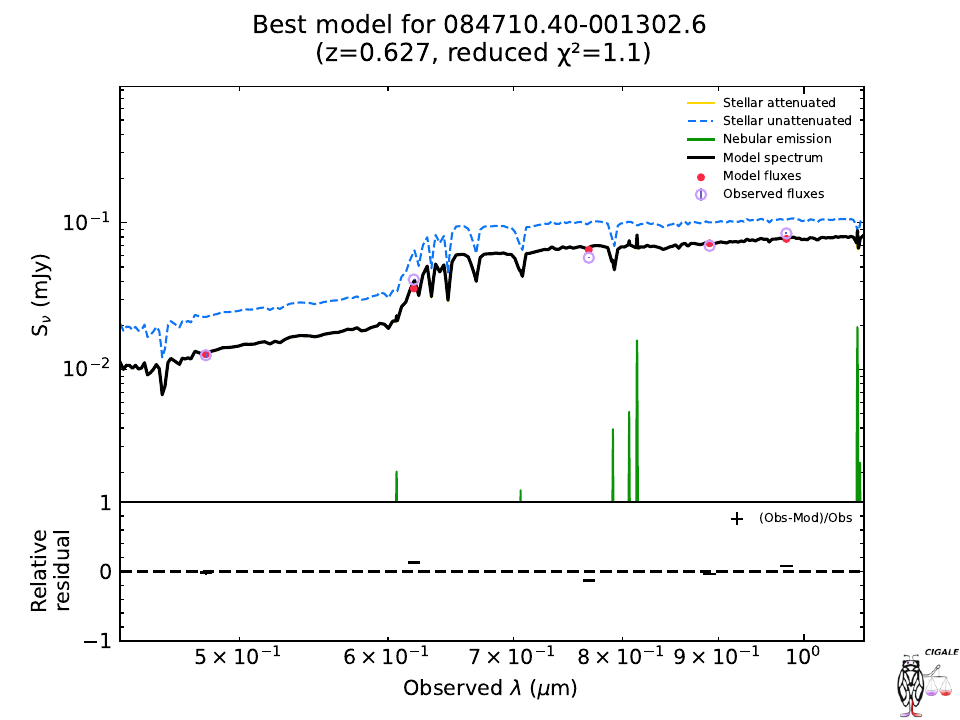}
\caption{SED fitting of SDSS J0847-0013 using \textsc{CIGALE}. The upper panel shows the data and the best-fit SED model. The lower panel shows the relative residual of the fitting, defined as $\text{(data-model)}/\text{data}$}
\label{fig:0847_cigale}
\end{figure*}

Since the primary focus of this work is on the molecular gas seen by CO emission, detailed optical measurements are not extensively presented in the main text. Nevertheless, for methodological completeness, we provide a concise summary of the key techniques employed to determine the optical properties of our dual quasars. As mentioned in Section \ref{subsec:selection}, the 2D image modelling tool \textsc{GaLight} is used to select out dual quasar candidates \citep{ding2022galight}. At the meantime, it is used to locate the center of the nuclei and estimate the magnitudes of the host galaxies after removing the point source contributions in all five bands \citep{silverman2020dual,tang2021optical}. Utilizing these magnitudes, we conduct spectral energy distribution (SED) fitting using the Code Investigating GALaxy Emission \citep[\textsc{CIGALE},][]{boquien2019cigale} tool. An example is presented in Figure~\ref{fig:0847_cigale} for SDSS J0847-0013. The measured magnitudes are converted to mJy and depicted as purple open circles with error bars. The red filled dots and the black curve represent the best-fit SED model. The model encompasses a delayed star formation history (SFH) and stellar population \citep{bruzual2003stellar, maraston2005evolutionary}, dust attenuation \citep{calzetti2000dust, charlot2000simple}, re-emission \citep{dale2014two}, and nebular emission \citep{inoue2011rest}, as indicated by the colored curves. The lower sub-panel displays the relative residual, defined as $\text{(data-model)}/\text{data}$. The resulting total stellar mass ($M_*$) is a relatively stable output of the fitting, listed in Table \ref{tab:optical} column (11).
\par

\begin{figure*}
\centering
\includegraphics[width=.9\linewidth]{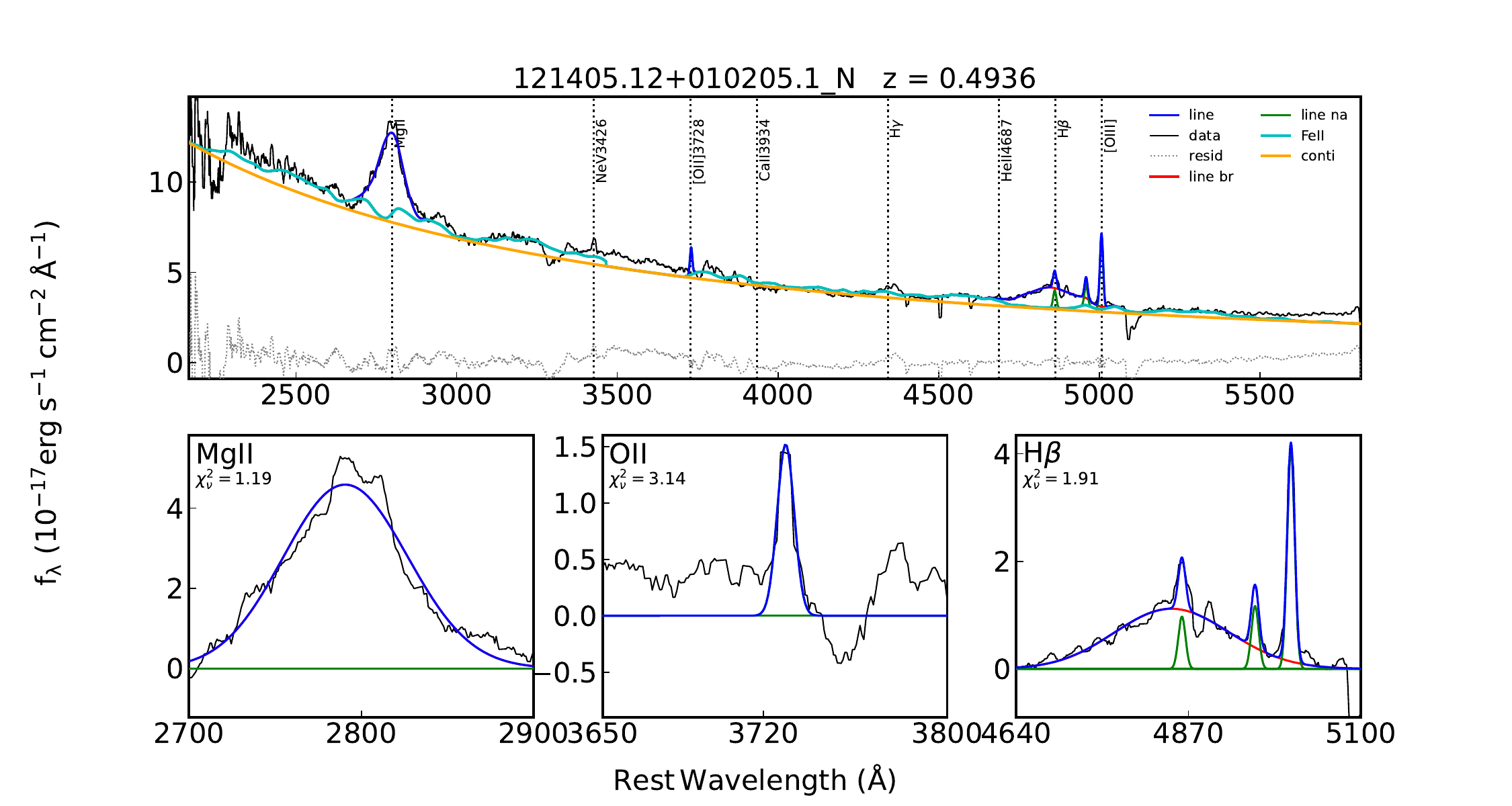}
\caption{Spectral fitting results of SDSS J1214+0102N with PyQSOFit. The upper panel shows the entire spectrum overlaid with the best-fit models, including the broad and narrow emission lines, iron emission, and continuum emission. The lower panel zooms into the regions of the specific emission lines.}
\label{fig:1214_pyqsofit}
\end{figure*}

On the spectroscopic front, BH properties are estimated using PyQSOFit \citep{guo2018pyqsofit}. Illustrated in Figure~\ref{fig:1214_pyqsofit} is an example for SDSS J1214+0102N. The upper panel plots the reduced Keck/LRIS data as a black curve, and the best-fit model categorized into broad and narrow emission lines, iron emissions, and continuum emission. The grey dotted curve represents the residual. The redshift is determined under the simultaneous fitting of all these components (Table~\ref{tab:optical} column 4).  The lower panels offer a closer look at specific emission line regions, each displaying the reduced $\chi^2$ value at the top left. The fitting results include the width of each emission line and monochromatic luminosity at specific wavelengths. Of particular interest are the broad emission lines, i.e., $\text{Mg \sc{ii}}$, H$\beta$, and H$\alpha$. Using their respective monochromatic luminosities, BH mass is estimated through the virial methods outlined in \cite{vestergaard2006determining, schulze2018fmos}. The outcomes are detailed in Table~\ref{tab:optical} columns (7)-(9). The monochromatic luminosity $L_{5100}$ (column 5) is employed to calculate the bolometric luminosity $L_{\mathrm{bol}}$ (column 6), allowing for the determination of the Eddington ratio $\lambda_{\rm Edd}$ (column 10).

\begin{table*}
\caption{Optical properties of each source of the dual quasars measured with PyQSOFit \citep{guo2018pyqsofit} and \textsc{CIGALE} \citep{boquien2019cigale}. The detailed methodology will be presented in a companion paper Tang et al. 2024 (in prep). \\
Columns (2-3): Central positions of the point sources estimated with our 2D image modeling tool \textsc{GaLight}. These positions correspond to the white crosses in Figure \ref{fig:poststamps}.
Column (5): Monochromatic luminosity at 5100 \text{\AA} ($L_{\rm 5100}$) measured from power law fitting to the continuum. The uncertainties of $L_{\rm 5100}$ is negligible with 200 MC samplers in PyQSOFit, as the shape of the power law is usually well-constrained.\\
Column (6): Bolometric luminosity ($L_{\rm bol}$) of the quasars measured with monochromatic luminosity $L_{\rm 5100}$ and bolometric correction factor BC$_{\rm 5100}=9.26$ according to \citet{shen2011catalog,richards2006spectral}. \\
Column (7-9): BH mass ($M_{\rm BH}$) measured with the viral method using broad $\text{Mg \sc{ii}}$ \citep{schulze2018fmos}, H$\beta$ \citep{vestergaard2006determining}, and H$\alpha$ lines \citep{schulze2018fmos}, respectively. According to \citet{shen2011catalog,shen2013mass,schulze2018fmos}, we assume the systematic error of the viral method to be 0.4 dex. the uncertainties reported here are a combination of the systematic error and the observational error from the spectra.\\
Column (10): Eddington ratio ($\lambda_{\rm Edd}$) measured as $L_{\rm bol}/L_{\rm Edd}$, where $L_{\rm Edd}=1.26\times10^{38}M_{\rm BH}$. $M_{\rm BH}$ takes the value following the priority H$\alpha > \mathrm{H}\beta > \mathrm{\text{Mg \sc{ii}}}$. \\
Column (11): Bayesian estimates of stellar mass from \textsc{CIGALE}. For J0847-0013 and J1416+0033, the two components share the same host galaxies, thus the \textsc{CIGALE} measurements are the same.\\}
\label{tab:optical}
\begin{tabular}{cccccccccccc}
\hline
Name & RA & Dec & z & $\log~L_{\rm 5100}$ & $\log~L_{\rm bol}$ & $\log~M_{\rm BH}^{\rm \text{Mg \sc{ii}}}$ & $\log~M_{\rm BH}^{\rm H\beta}$ & $\log~M_{\rm BH}^{\rm H\alpha}$ & $\log~\lambda_{\rm Edd}$ & $\log~M_*$\\
 & (hh:mm:ss) & (dd.mm.ss) &  & ($\mathrm{erg~s^{-1}}$) & ($\mathrm{erg~s^{-1}}$) & ($\mathrm{M_{\odot}}$) ($\mathrm{M_{\odot}}$) & ($\mathrm{M_{\odot}}$) &  & ($\mathrm{M_{\odot}}$)\\
(1) & (2) & (3) & (4) & (5) & (6) & (7) & (8) & (9) & (10) & (11)\\
\hline
J0847-0013N & 08:47:10.402 & -00:13:02.460 & 0.6256 & 44.53 & 45.49 & $9.02\pm0.40$ & -- & -- & -1.65 & $10.90\pm0.21$ \\ 
J0847-0013S & 08:47:10.440 & -00:13:03.288 & 0.6269 & 44.39 & 45.26 & $8.72\pm0.40$ & -- & -- & -1.57 & $10.90\pm0.21$ \\ 
J1214+0102N & 12:14:05.110 & 01:02:07.188 & 0.4936 & 44.11 & 45.04 & $8.83\pm0.40$ & $9.06\pm0.40$ & -- & -2.13 & $10.52\pm0.17$ \\ 
J1214+0102S & 12:14:05.134 & 01:02:05.028 & 0.4916 & 43.82 & 44.85 & $8.22\pm0.40$ & $8.71\pm0.40$ & -- & -1.98 & $10.65\pm0.12$ \\ 
J1416+0033W & 14:16:37.418 & 00:33:52.416 & 0.4331 & 43.70 & 44.50 & $8.85\pm0.40$ & $9.39\pm0.40$ & $9.11\pm0.59$ & -2.72 & $10.84\pm0.15$ \\ 
J1416+0033E & 14:16:37.459 & 00:33:52.200 & 0.4328 & 42.94 & 44.28 & $7.85\pm0.40$ & $8.60\pm0.40$ & $8.64\pm0.43$ & -2.48 & $10.84\pm0.15$ \\ 
J2209+0045N & 22:09:06.912 & 00:45:43.848 & 0.4458 & 43.70 & 44.66 & -- & $8.14\pm0.40$ & $8.01\pm0.40$ & -1.46 & $10.07\pm0.08$ \\ 
J2209+0045S & 22:09:06.900 & 00:45:42.228 & 0.4457 & 43.10 & 44.07 & -- & $7.82\pm0.40$ & $7.50\pm0.40$ & -1.54 & $10.57\pm0.13$ \\ 
J2337+0056N & 23:37:13.694 & 00:56:12.048 & 0.7083 & 43.92 & 44.89 & -- & $7.90\pm0.40$ & -- & -1.12 & $10.05\pm0.25$ \\ 
J2337+0056S & 23:37:13.673 & 00:56:10.752 & 0.7089 & 44.10 & 45.06 & -- & $8.16\pm0.83$ & -- & -1.22 & $10.94\pm0.20$ \\ 
\hline
\end{tabular}
\end{table*}

\bsp	
\label{lastpage}
\end{document}